\documentclass[twocolumn]{IEEEtran}

\usepackage{epsf}
\usepackage{graphicx}
\DeclareGraphicsExtensions{.png}
\usepackage{caption}
\usepackage{subcaption}
\usepackage{xcolor}
\usepackage{amsmath}
\usepackage{amsfonts}
\usepackage{amssymb}
\usepackage{cuted}
\usepackage{float}
\usepackage[utf8]{inputenc}
\usepackage[OT2, T1]{fontenc}
\usepackage[french,english]{babel}
\newcommand\textcyr[1]{{\fontencoding{OT2}\fontfamily{wncyr}\selectfont #1}}

\newcommand{{\calB}}{{\cal B}}

\newcommand{{\calC}}{{\cal C}}

\newcommand{{\calD}}{{\cal D}}

\newcommand{{\calE}}{{\cal E}}

\newcommand{{\calH}}{{\cal H}}

\newcommand{{\calJ}}{{\cal J}}

\newcommand{{\calM}}{{\cal M}}

\newcommand{{\calQ}}{{\cal Q}}

\newcommand{{\calP}}{{\cal P}}

\newcommand{{\calS}}{{\cal S}}

\newcommand{{\calU}}{{\cal U}}

\newcommand{{\calW}}{{\cal W}}

\newcommand{\sinc}{{\rm sinc}}

\newcommand{\bea}{\begin{eqnarray}}

\newcommand{\eea}{\end{eqnarray}}

\newcommand{\rmd}{{\rm d}}

\newcommand{\rmeff}{{\rm eff}}

\newcommand{\rmerf}{{\rm erf}}

\newcommand{\rmerfi}{{\rm erfi}}

\newcommand{\rmj}{{\rm j}}

\newcommand{\rmmax}{{\rm max}}

\newcommand{\rmth}{{\rm th}}

\newcommand{\rmRe}{{\rm Re}}

\newcommand{\rmIm}{{\rm Im}}

\begin{document}
\title{Excess Power, Energy and Intensity of
Stochastic Fields in Quasi-Static and Dynamic Environments}

\author{\IEEEauthorblockN{Luk R. Arnaut} %{\it Senior Member IEEE}
}

\maketitle

\begin{abstract}
The excess power, energy and intensity of a random electromagnetic field above a high threshold level are characterized based on a Slepian--Kac model for upcrossings. 
For quasi-static fields, the probability distribution of the excess intensity in its regression approximation evolves from $\chi^2_3$ to $\chi^2_2$ when the threshold level increases.
The excursion area associated with excess energy exhibits a chi-cubed ($\chi^3_2$) distribution above asymptotically high thresholds, where excursions are parabolic.
For dynamic fields, the dependence of the electrical and environmental modulations of the excess power on the hybrid modulation index and threshold level are established.  
The normalized effective power relative to the quasi-static power increases non-monotonically when this index increases.
The mean and standard deviation of the dynamic excess power are obtained in closed form 
and validated by Monte Carlo simulation.
\end{abstract}

{\bf \small {\it Index Terms} -- 
excursion area, extreme electromagnetics, immunity testing, peaks over threshold, regression, reverberation chambers, Slepian model, threshold level crossings.}

\section{Introduction}
Random fluctuations of electromagnetic (EM) fields and currents have been studied for over a century, typically as noise phenomena in a  static or quasi-static EM environment (EME).
Dynamic EMEs introduce additional field fluctuations superimposed onto heterodyne signal modulation of a purely electrical origin. In this respect, mode-tuned and mode-stirred reverberation chambers (MT/MSRCs) serve as generators of quasi-static and dynamic spatial fields in the multipath propagation of time-harmonic or modulated signal fields. 

In modern wireless communications and mobile computing, the signalling systems and devices typically involve an exceedingly large number of independent states (degrees of freedom $N$), e.g., in wideband multi-user communications for multimedia and high-mobility or intelligent transport applications \cite{reib2019}. 
This results in complex systems that must be verified and operated in real time under rapidly varying conditions and continuous state transitions, with minimum latency and uncertainty. 
In this context, the generation \cite{drika2020} and time-domain measurement \cite{krug2005}--\cite{arnajitt} of ultra-short pulses and peaks are active areas of research. 
One important parameter is the peak to average power ratio (PAPR), which is highly sensitive to $N$.

Specific to testing of complex systems in MT/MSRCs, mechanical stirring may no longer be feasible to provide sufficiently short cycle times per test state. 
Rapid electronic multistirring offers much shorter test cycles, e.g., electronic source switching or scanning. 
In conjunction with increasing rates and shortened events in dynamic EMEs, quasi-static concepts such as effective power may need to be reconsidered at microwave and millimeter-wave frequencies. 
  
To this end, a statistical characterization of excursions for energy and power above a constant threshold level was made in \cite{arnaexcur} for mono- and multi-stirred or -scanned MSRCs.
There, the focus was on quasi-static operation -- i.e., slow variations of the field envelope relative to the rate of oscillation of the excitation field at its carrier frequency -- and on  temporal characteristics, i.e., location, duration and rate of threshold crossings and excursions.
Such an analysis complements traditional EM characterization based on field amplitudes. 
The rationale is
that EM stress and failure of equipment under test may not only be governed by the instantaneous field amplitude or intensity, i.e., the ``vertical'' level, but also by their temporal, i.e., ``horizontal''  sustainment, interruptions, durations and rates for high levels above a critical threshold.

\begin{figure}[!ht] 
\begin{center} \begin{tabular}{c}
\vspace{-3.9cm}
\hspace{-0.7cm}
%FIGURE 1 IN diagram_ExcessArea_v2.m
\includegraphics[scale=0.48]{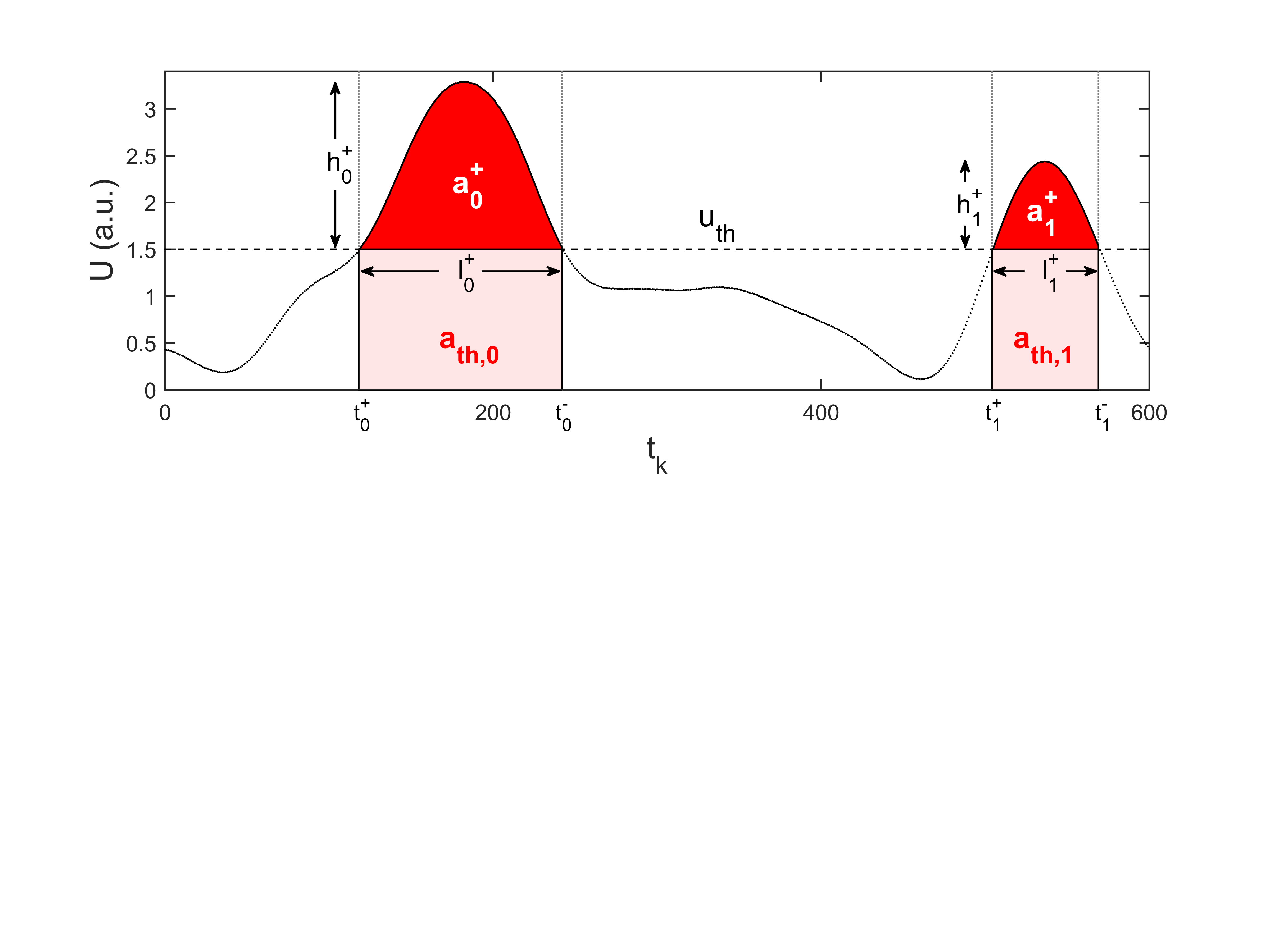}\ \\
\end{tabular} \end{center}
{\small 
\caption{\label{fig:diagram_Apls}
\small Two realizations of positive excursions for $U(t)$ (in arbitrary units), with threshold up- and downcrossing times $t^+_i$ and $t^-_i$, excursion length $\ell^+_i$, height $h^+_i$, area $a^+_i$, threshold bounded area $a_{\rmth, i}$ and total area $a_{i}=a^+_i+a_{\rmth, i}$ within time interval $[t^+_i,t^-_i]$ ($i=0,1$). 
}
}
\end{figure}

In this paper, horizontal and vertical metrics are combined into a comprehensive time-domain characterization. The focus is on the random height and area of positive excursions above a relatively high threshold (Fig. \ref{fig:diagram_Apls}), from which probability density functions (PDFs) of excess energy and power are deduced. To this end, the statistics of the energy and its derivative \cite{arnalocavg} are fundamental. 
Although only a single mechanism of variation for the EME will be considered, e.g., one-dimensional (1-D) monostirring in a MSRC, an extension to $d$  dimensions of randomization where $d$ stir actions are represented by multiple time, space and/or state dimensions \cite{arnaexcur} is possible. 
The emphasis is on scalar electric power defined by the electric energy per unit of time, as opposed to vector EM power density defined by the Poynting vector.
Random variables are denoted by uppercase letters, while their values and constants are written in corresponding lowercase. An $\exp(\rmj \omega t)$ time dependence is assumed.
%**************FIGURE 01****************
%**************FIGURE 01****************

\section{Slepian--Kac Model for Upcrossings}
A Slepian (or Slepian--Kac) model (SKM) \cite{kac1959}, \cite{slep1963} represents a decomposition of a noise process, conditioned on crossing a predetermined threshold level, into a short-term quasi-deterministic component and a longer-term nonstationary random component. The former can be conceived as a regression approximation \cite{lind1982} in the vicinity of an upcrossing, downcrossing, local extremum or other mark for each positive excursion (exceedance) by an EM quantity or its temporal or spatial derivative.
The SKM is implicit in the modelling of the excursions' duration or spatial extent (length or area) \cite{arnaexcur}, as the footprint of a $(d+1)$-D excursion surface or volume projected onto a $d$-D threshold contour or surface, respectively.
In the present work, a SKM will be developed for the 1-D excess height and 2-D excess area, with the associated instantaneous and integrated energy (or power), respectively. To establish notions and notations, the basic SKM conditioned on level upcrossings is briefly summarized first.

\subsection{Gauss Normal Field or Current Process\label{sec:SKMfield}}
Let $X(t)$ be a sufficiently smooth zero-mean real-valued Gaussian random EM field or current, assumed to be stationary and ergodic to enable a probabilistic characterization of $X(t)$ 
from a single long-run realization (sample path, sweep) as a time series.
Assume that pairs of up- and downward crossings of a chosen arbitrary threshold level $x_{\rmth}$ by $X(t)$ occur at respective times $t^+_i$ and $t^-_i$ ($i=0,1,2,\ldots$) bounding the $i$th positive excursion $X^+(t)$.
First, consider a single excursion starting at $t^+=0$, i.e., $X(0)=x_{\rmth}$.
The second-order SKM for $X^+(t)$ applicable to this $X(t\geq 0)$ is \cite{slep1963}
\begin{align}
X^+({t|{x_{\rmth}}}) = 
\left [ \frac{r_X(t)}{\lambda_0} X(0) - \frac{\dot{r}_X(t)}{{\lambda_2}} \dot{X}(0) \right ] + K(t).
\label{eq:defSlepianGauss_up}
\end{align}
Here, $\dot{r}_X(t) 
= - \langle{\dot{X}(t^+)X(t^+ + t)}\rangle$ is the time derivative of the ensemble (and, by ergodicity, temporal) autocovariance function\footnote{Autocovariance functions containing an expansion term in $|t|^3$ give rise to differences in $f_{L^+}(\ell^+\rightarrow 0)$, among other effects \cite[sec. 3]{slep1963}.}
$r_X(t)$ of $X(t)$, expandable for $t \ll t^- -t^+$ as 
\begin{align}
r_{X}(t) &= \langle{X(0)X(t)} \rangle
= \lambda_0 - \frac{\lambda_2}{2!} t^2 + \frac{\lambda_4}{4!} t^4 + {\cal O}(t^6)
\label{eq:acfX}
\end{align}
where $\lambda_0 =\sigma^2_{X}$, $\lambda_2 =\sigma^2_{\dot{X}}$ and $\lambda_4 =\sigma^2_{\ddot{X}}$ are the finite zeroth, second and fourth spectral moments of $X(t)$, respectively; 
$\dot{X}(0) \stackrel{\Delta}{=} Z /\sqrt{\lambda_2} >0$ 
is the random slope of $X(t)$ at $t^+=0$, which in the case of so-called horizontal windowing \cite[eq. (2.1)]{kac1959} exhibits a scaled Rayleigh ($\chi_2$) PDF where 
\begin{align}
f_{Z}(z) = z \exp \left ( - z^2/2 \right ),\,\,\,\,\,\,z>0
\label{eq:fZ}
\end{align} 
with sample value $z_i$ at $t^+_i$ for the $i$th excursion;
and
$K(t)$ is a nonstationary residual Gaussian stochastic field \cite{slep1963} with time average $\overline{K(t)} = 0$ and inhomogeneous autocovariance function
\begin{align}
r_K(t_1,t_2) &\equiv
\overline{ K(t_1) K(t_2) } 
= \overline{ X(t_1) X(t_2) \, | \, \left ( X(t^+), \dot{X}(t^+) \right ) }\nonumber\\
&= r_{X}(t_1-t_2) - r_{X}(t_1) r_{X}(t_2) / \lambda_0 \nonumber\\
&~~ - \dot{r}_{X}(t_1) \dot{r}_{X}(t_2) / \lambda_2,~~~~~t_1,t_2\in [t^+,t^-].
\label{eq:autocov_K}
\end{align}
The $\dot{X}(0)$ and $K(t)$ in (\ref{eq:defSlepianGauss_up}) are mutually independent.
The bracketed sum in (\ref{eq:defSlepianGauss_up}) represents a regression approximation $\tilde{X}^+(t|x_{\rmth}) \stackrel{\Delta}{=} \alpha_X(t) + \beta_X(t) Z$ for $X^+(t|x_{\rmth})$
as a linear combination of two orthogonal basis functions $X(0)$ and $\dot{X}(0)$ \cite{blach1993}, plus a residual fluctuation $K(t)$ for the actual excursion. 
For a narrowband $X(t)$, where $|r_K(t,t)| \ll |r_X(t)| $ when $ t \ll t^- - t^+$ \cite[sec. 1]{slep1963}, the regression $\tilde{X}^+(t|x_{\rmth})$ dominates $K(t)$. 
For larger $t$, the magnitude of $K(t)$ may become comparable to, or exceed $\tilde{X}^+(t|x_{\rmth})$.
%and also self-uncorrelated at the same ($\langle X(t) X^\prime(t) \rangle =0)$; 
Higher-order extensions of the SKM (\ref{eq:defSlepianGauss_up}), conditioned on additional knowledge of higher-order derivatives at $t^+$ or multiple first-order derivatives at $t \not =  t^+$, are also possible \cite[sec. 2]{slep1963}, \cite[sec. VI]{blach1988}.

For economy of notation, we further denote the (dimensionless) normalized quantities $\lambda^\prime_{j} \stackrel{\Delta}{=} {\lambda_{j} } / { \lambda_0 }$ and define
\begin{align}
x^\prime(t) \stackrel{\Delta}{=} \frac{x(t)}{\sigma_X},\,
u^\prime(t) \stackrel{\Delta}{=} \frac{u(t)}{\sigma^2_X},\,
r^\prime_X(t) \stackrel{\Delta}{=} 
\frac{r_X(t)}{\sigma^2_X},\,
r^\prime_U(t) \stackrel{\Delta}{=} 
\frac{r_U(t)}{\sigma^4_X}
.
\label{eq:def_norm}
\end{align}
With these normalizations, $r^\prime_X(t)$ is the autocorrelation function (ACF) of $X(t)$. In particular, from (\ref{eq:autocov_K}), the variance function of $K(t)$ becomes $r^\prime_K(t,t) = \sigma^2_{K}(t)/\sigma^2_{X}$, i.e.,
\begin{align}
r^\prime_K(t,t) = 1 - [r^\prime_X(t)]^2 - \left [ {\dot{r}^\prime_X(t)} / {\sqrt{\lambda^\prime_2}} \right ]^2.
\label{eq:acf_K}
\end{align} 

\begin{figure}[!ht] 
\begin{center} \begin{tabular}{c}
\vspace{-3.7cm}
\hspace{-0.9cm}
%FIGURE 100 IN acf_v1.m
\includegraphics[scale=0.49]{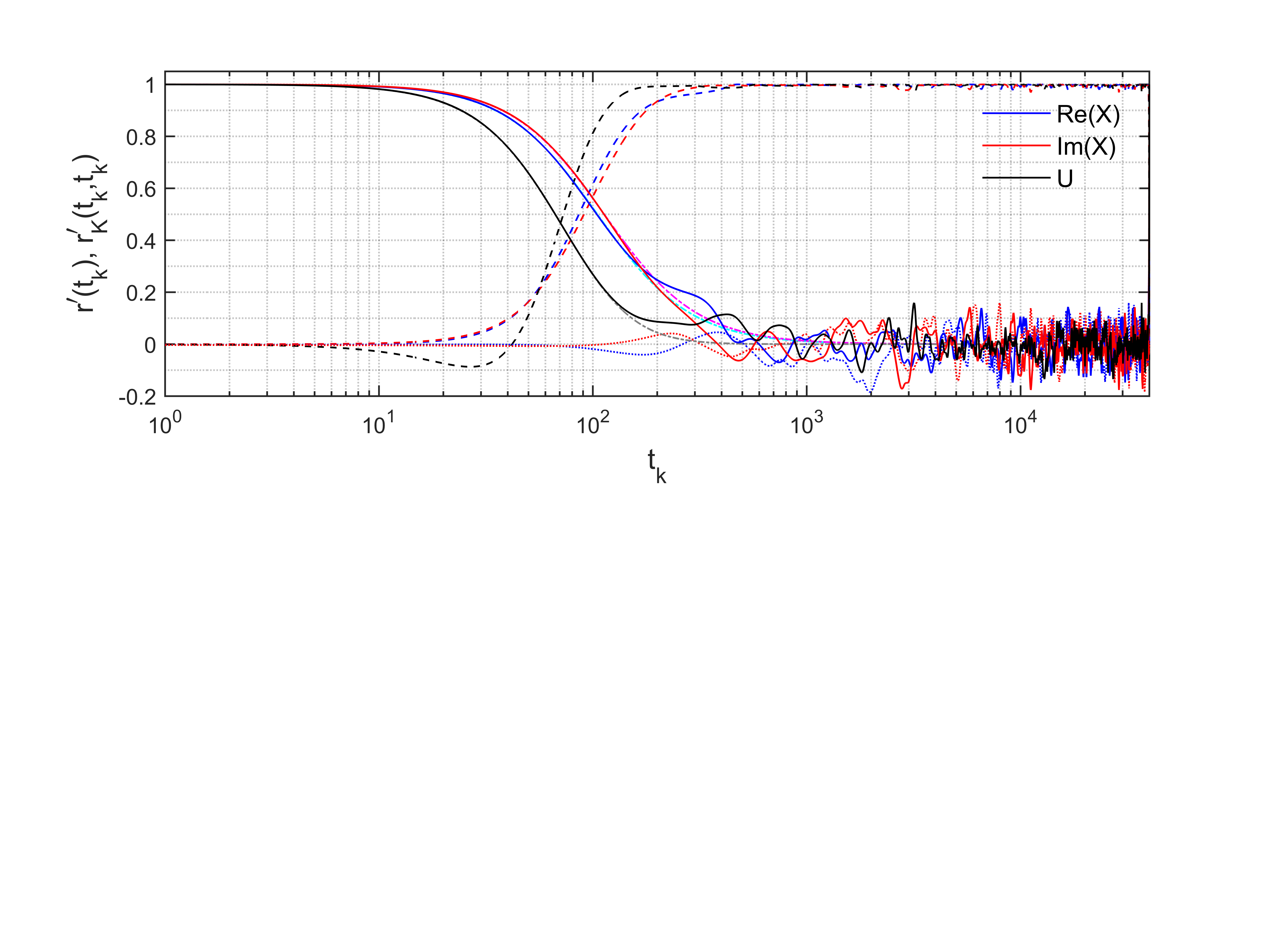}\ \\
(a)\\
\vspace{-4.0cm}
\hspace{-0.9cm}
%FIGURE 100 IN acf_v1.m
\includegraphics[scale=0.49]{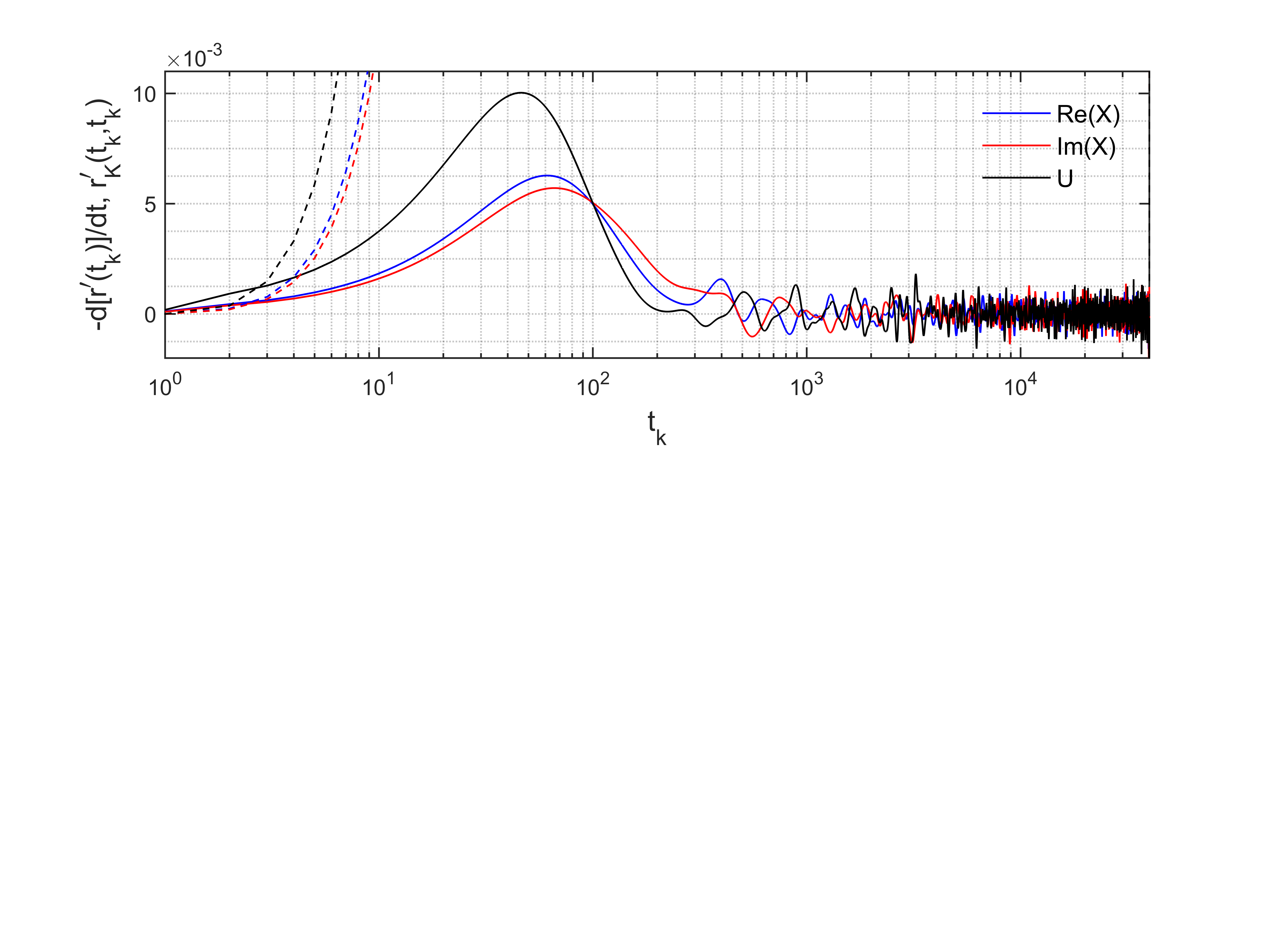}\ \\
(b)\\
\end{tabular} \end{center}
\caption{\label{fig:acfXUK}
{\small 
(a)-(b)
Autocorrelation functions: (a) $r^\prime_{X}(t_k)$, $r^\prime_U(t_k)/4$ (solid) and 
(b) $-\dot{r}^\prime_X(t_k)$, $-\dot{r}^\prime_U(t_k)/4$ 
(solid) 
of measured ${\rmRe}[X(t)]$ (blue), ${\rmIm}[X(t)]$ (red) and $U(t)$ (black) for $n=2$, with comparison to
$r^\prime_K(t_k,t_k)$ for respective $K(t)$ (dashed)
and to theoretical ACF models (dash dotted)
$\rho_{{\rmRe}(X)}(t_k) = 1/[1+(t_k/105\Delta t_k)^2]$ (cyan), 
$\rho_{{\rmIm}(X)}(t_k) = 1/[1+(t_k/113\Delta t_k)^2]$ (magenta), and 
$\rho_{U}(t_k) = 1/[1+(t_k/103.5\Delta t_k)^2]^2$ (grey).
(a) I/Q cross-correlation functions (dotted): $r^\prime_{{\rmRe}(X),{\rmIm}(X)}(t_k)$ (blue) and $r^\prime_{{\rmIm}(X),{\rmRe}(X)}(t_k)$ (red).
}
}
\end{figure}

Fig. \ref{fig:acfXUK} shows $r^\prime_X(t_k)$, $-\dot{r}^\prime_X(t_k)$ and $r^\prime_K(t_k,t_k)$ for MSRC data of measured electric field components ${\rmRe}[X(t_k)]$ and ${\rmIm}[X(t_k)]$, based on experimental data detailed in Sec. \ref{sec:exp} and $t^+_k\stackrel{\Delta}{=}0$. 
It is seen that, immediately after upcrossings ($t_{k} \ll 100\,\Delta t_k$), $r^\prime_X(t_k)$ 
outweighs $r^\prime_K(t_k)$ while for $t_{k} \gg 300\,\Delta t_k$, the opposite holds. 
Since $r^\prime_X(t_k)$ and $r^\prime_K(t_k)$ cross at a relatively high level, both contributions are significant at intermediate times. 
From Fig. \ref{fig:acfXUK}(b), contributions by $-\dot{r}^\prime_X(t_k)$ are relatively small compared to $r^\prime_K(t_k)$.
Thus, the regression $\tilde{X}(t|x_{\rmth})$ dominates the short-term field, i.e., it governs the functional and statistical dependence within the correlation length $t_c$, {\it a fortiori} within the excursion length $\ell^+\stackrel{\Delta}{=}t^+ - t^- < t_c$. 
On the other hand, sufficiently long after an upcrossing at $t^+$, the latent field $K(t)$ tends to become uncorrelated in the long term, well known from coarsely sampled or unconditioned processes. The conditioned field thus becomes increasingly 
less constrained by the earlier field and its derivative at $t^+$ and becomes asymptotically stationary.

Fig. \ref{fig:acfXUK}(a) also shows instantaneous cross-correlation functions between ${\rmRe}[X(t_k)]$ and ${\rmIm}[X(t_k)]$. Their low magnitude smaller than 0.1 across the entire time span justifies adopting a SKM for independent processes \cite{lind1989} in Sec. \ref{sec:SKMenergy}.
%**************FIGURE 02****************
%**************FIGURE 02****************

\subsection{$\chi^2_n$ Field Intensity or Current Intensity Process\label{sec:SKMenergy}}
Consider the $\chi^2_n$ distributed $U(t)
=\sum^n_{j=1} X^2_j(t)$ for $n$ independent and identically distributed (i.i.d.) real $X_j(t)$ as defined in Sec. \ref{sec:SKMfield}, e.g., a set of $n$ I/Q 
Cartesian components of the EM field. 
It follows that $u_{(\rmth)}/\sigma^2_{X_j} = n \, u_{(\rmth)}/\sigma_U$ and $r_U(t) = 2n \, r^2_{X_j}(t)$.
The SKM for $U^+(t\geq 0)$ conditioned on an upcrossing of $u_{\rmth}=U(0)$ at $t^+=0$ is \cite[Thm. 2]{lind1989}  
\begin{align}
{U^+}(t | u_{\rmth}) = \lambda_0 \sum^{n}_{j=1} 
\left [ \epsilon_j \alpha(t) + \beta(t) W_j + K_j (t) 
\right ]^2 
\label{eq:defSlepianChiSq_up}
\end{align}
with $\epsilon_1=1$, $\epsilon_{j\not=1}=0$, and regression coefficient functions
\begin{align}
\alpha(t) \stackrel{\Delta}{=} \sqrt{\frac{u_{\rmth} r_{U}(t)}{2n \lambda^3_0} }, \,\,\,\,\,\,
\beta(t) \stackrel{\Delta}{=} \frac{- \dot{r}_{U}(t)}{\sqrt{8n \lambda_0 \lambda_2 r_U(t)}}
\label{eq:alphabetaU}
\end{align}
where $W_1\equiv Z$ has a standard Rayleigh $\chi_2$ PDF; $W_2,\ldots, W_n$ are mutually independent standard Gauss normal and also independent of $W_1$; all $K_j(t)$ are independent nonstationary standard Gauss normal processes with zero mean ($\overline{K_j(t)}=0$) and common autocovariance function (\ref{eq:autocov_K}).
For $t - t^+ \ll t_c$, the autocovariance of $U(t)$ can be expanded as
\begin{align}
r_{U}(t) 
&= 
2n \left [ \lambda^2_0 - \lambda_0 \lambda_2 t^2  + \frac{3\lambda^2_2+\lambda_0\lambda_4}{12} t^4 + {\cal O}(t^6) \right ]
.
\label{eq:acfU}
\end{align}
Examples of non-Gaussian $r_U(t)$ 
%for MSRCs 
were given in \cite{arnalocavg}.

In the remainder, the focus is on a single Cartesian circular complex field $X(t)={\rmRe}[X(t)]+\rmj {\rmIm}[X(t)]$ associated with $U(t)$, i.e., $n=2$. 
From (\ref{eq:defSlepianChiSq_up}), the normalized mean and variance of $U^+$ then follow after calculation as
\begin{align}
\langle {U^+}^\prime(t|u^\prime_{\rmth}) \rangle
&= \alpha^2 + \sqrt{2\pi} \, \alpha\beta + 3 \beta^2 + 2 \langle K^2 \rangle
\label{eq:mean_Upls}
\end{align}
\begin{align}
{\sigma^\prime}^2_{{U^+}}(t|u^\prime_{\rmth}) &=
(8-2\pi)\alpha^2\beta^2 + 2 \sqrt{2\pi} \, \alpha \beta^3 + 6 \beta^4 \nonumber\\
&~~ + \left ( 4 \alpha^2 + 4\sqrt{2\pi} \, \alpha\beta + 12 \beta^2 
+ 4 \langle K^2 \rangle \right ) \langle K^2  \rangle
\label{eq:var_Upls}
\end{align}
where $\langle K^2(t) \rangle$ is the normalized variance function (\ref{eq:acf_K}) for any $K_j(t)$ in (\ref{eq:defSlepianChiSq_up}), which can be expressed equivalently as
\begin{align}
\langle K^2(t) \rangle
=
1 - [\alpha^2(t) / u^\prime_{\rmth}] - \beta^2(t).
\end{align}

\begin{figure}[!ht] 
\begin{center} \begin{tabular}{c}
\vspace{-3.7cm}
\hspace{-0.8cm}
%FIG 2000 in acf_v5.m
\includegraphics[scale=0.48]{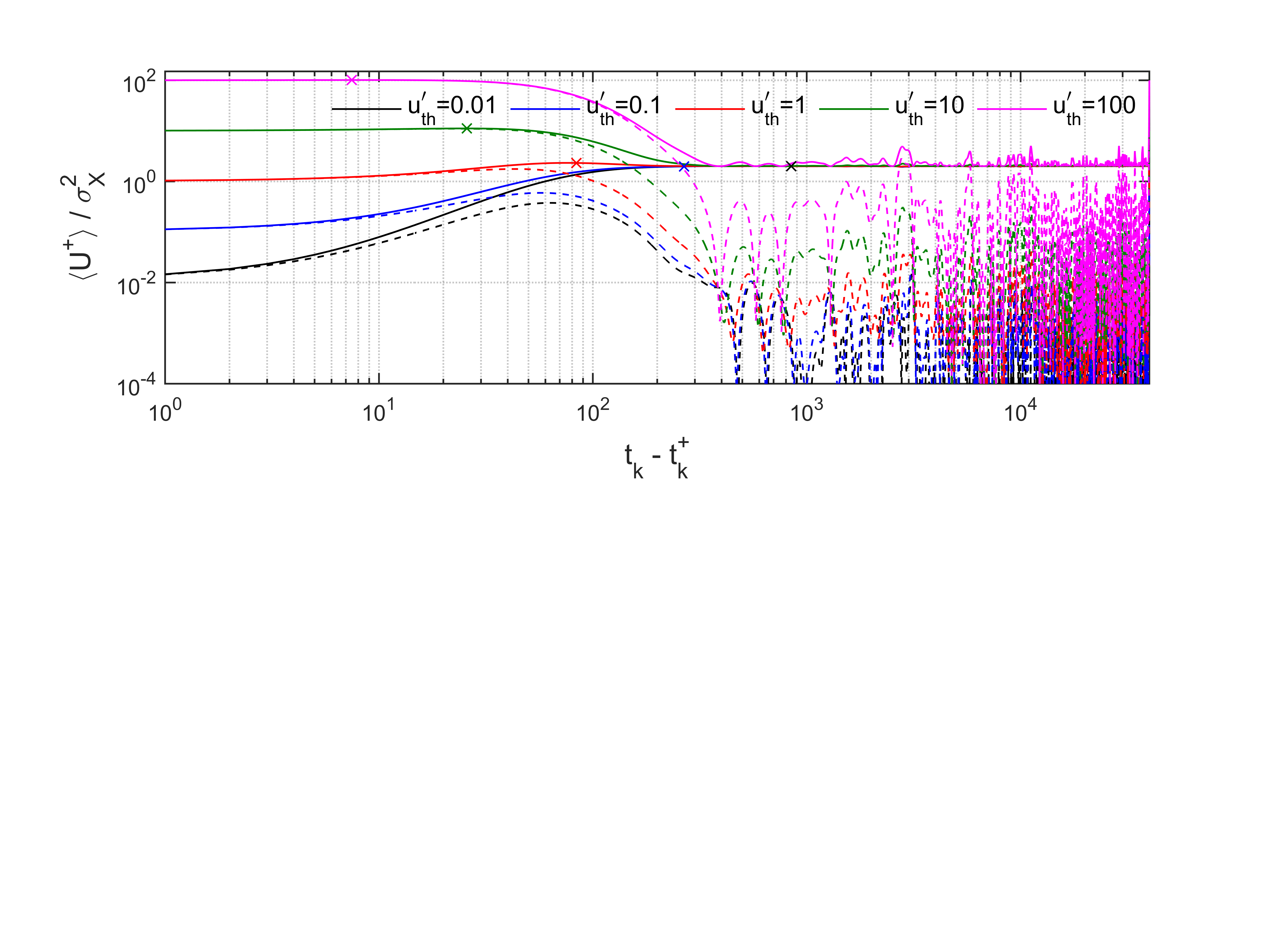}\ \\
(a)\\
\vspace{-3.7cm}
\hspace{-0.8cm}
%FIG 2009 in acf_v5.m
\includegraphics[scale=0.48]{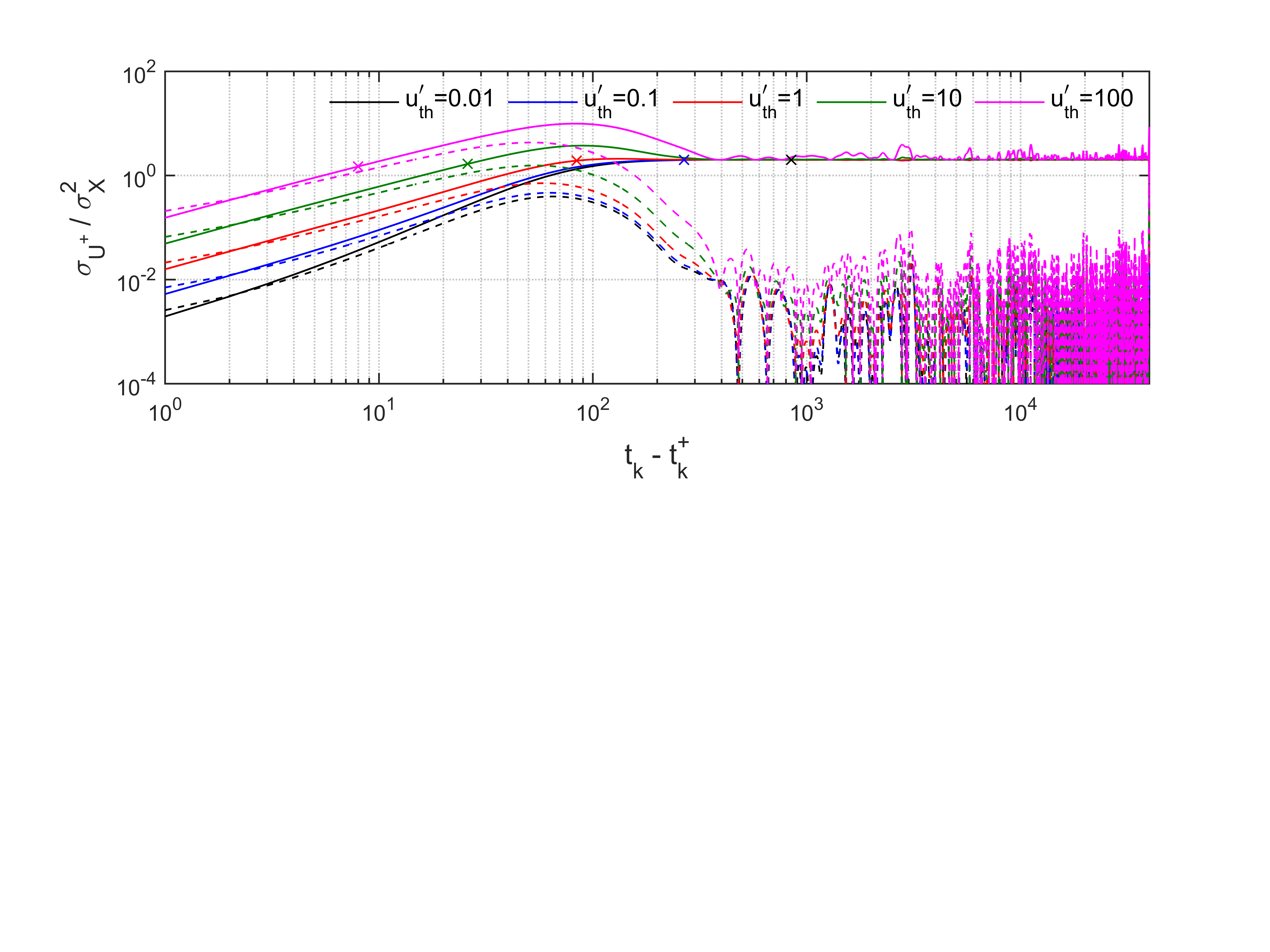}\ \\
(b)\\
\end{tabular} \end{center}
{\small 
\caption{\label{fig:Slepian_musigma}
\small
(a) Normalized mean $\langle {U^+}^\prime \rangle$ and (b) normalized standard deviation $\sigma_{{U^+}^\prime}$ (solid) of intensity $U^+$ with regression approximations $\langle \tilde{U}^{+^\prime} \rangle$ and $\sigma_{\tilde{U}^{+^\prime}}$ (dashed) for $n=2$ as a function of $t_k-t^+_k$ (in units $\Delta t_k$), after upcrossing of level $u^\prime_{\rmth}$ at $t^+_k$, based on experimental $r^\prime_U(t_k)$.
Cross symbols represent values at $t^*_k-t^+_k=\langle L^+ (u^\prime_{\rmth}) \rangle / 2$ situated near the expected local maxima of $U^+(t_k-t^+_k)$.
}
}
\end{figure}

Fig. \ref{fig:Slepian_musigma} shows the measured normalized mean and standard deviation of $U^+$ as a function of latency $t_k-t^+_k$ for selected values of $u^\prime_{\rmth}$, based on the empirical ACF $r_U(t_k-t^+_k)/\sigma^2_U$ that closely matches\footnote{For the present data set, (\ref{eq:acfU_model}) was found to outperform both Gaussian and $\sinc^2$-type models of $r_U(t_k-t^+_k)/\sigma^2_U$ for $0\leq t_k - t^+_k < 2t_c$.} 
 the model (cf. Fig. \ref{fig:acfXUK} and Sec. \ref{sec:exp})
\begin{align}
\rho_U(t)  = \{1+[t/(2 t_c)]^2\}^{-2},\,\,\,\,\, 2t_c \simeq 103.5\,\Delta t.
\label{eq:acfU_model}
\end{align} 
The results show that for $t_k - t^+_k \ll 2t_c$ or $\gg 2t_c$, the first or last term inside the brackets in (\ref{eq:defSlepianChiSq_up}) dominates, respectively. 
For the former, the regression approximation inside excursions is justified by its close correspondence. 
For $t_k-t^+_k\rightarrow +\infty$, both $\langle{U^+}\rangle$ and $\sigma_{U^+}$ converge to $\langle U \rangle = \sigma_{U} = 2\sigma^2_X$ for $\chi^2_2$, while their regression approximations approach zero, on average. 

Thus, a threshold based characterization has merit in the sense that the uncertainty of $U(t)$ during an excursion is significantly lower than in conventional, i.e., unconditioned characterization of $U(t)$ without thresholding.
%**************FIGURE 03****************
%FIG ??? in acf_v5.m
%**************FIGURE 03****************

\section{Excess Duration, Intensity and Energy}
\subsection{Excursion Length (Excess Duration)}
A PDF of the excursion size (hyperlength) $L^+_d$ along $d$ randomization dimensions was obtained in \cite{arnaexcur}. 
For $d=1$ (cf. Fig. \ref{fig:diagram_Apls}),
the excursion length $L^+(u_{\rmth}) 
\stackrel{{\cal L}}{\rightarrow} 
2 {W_1} / \sqrt{{\lambda^\prime_2 u^\prime_{\rmth}}}$ exhibits asymptotically a Rayleigh PDF 
\begin{align}
f_{L^+}(\ell^+)
\rightarrow
\frac{\lambda^\prime_2 u^\prime_{\rmth}}{4} \ell^+ \exp \left [ - \frac{\lambda^\prime_2 u^\prime_{\rmth}}{8} (\ell^+)^2 \right ]
\label{eq:PDF_Lpls_1}
\end{align}
with mean and standard deviation given by
\begin{align}
\langle L^+ \rangle = \sqrt{\frac{2\pi}{\lambda^\prime_2 \, u^\prime_{\rmth}}},~~ 
\sigma_{L^+} = \sqrt{\frac{8-2\pi}{\lambda^\prime_2 \, u^\prime_{\rmth}}}
.
\label{eq:avgstd_Lpls_1}
\end{align}

\subsection{Excursion Height (Excess Intensity)}
\subsubsection{Independently Sampled Field\label{eq:excursheight_discrete}}
It is well known \cite{balk1974}
that, for i.i.d. samples $X_i$ from a general $X(t)$ with PDF $f_X(x)$ and $x^\prime_{\rmth} \rightarrow +\infty$, the exceedances 
$H_{X,i} \stackrel{\Delta}{=} X_i-x_\rmth \geq 0 $ above $x_{\rmth}$ (also known as peaks over threshold (POT)) exhibit a distribution that belongs to one of three subclasses constituting the generalized Pareto distribution (GPD) given by
\begin{align}
G_X(x) \stackrel{\Delta}{=} F^*_{H_X}(h_X) = \lim_{\theta \rightarrow \gamma} \left [ 1 - \left ( 1 + \theta \frac{h_X-\mu_{H_X}}{\sigma_{H_X}} \right )^{-\frac{1}{\theta}} \right ].
\label{eq:GPD}
\end{align} 
The GPD is a limit distribution, 
in the sense that it applies to $X_i$ approaching the right end point $x_o$ of the support of $f_X(x)$ as a limit; here, $x_o \rightarrow +\infty$ for the present case of ideal unbounded fields and energy.
In (\ref{eq:GPD}), $\sigma_{H_X}>0$, $\mu_{H_X}$, and $\gamma$ are scale, location, and shape parameters, with $\gamma<0$, $=0$, and $>0$ yielding beta, Gumbel, and simple Pareto distributions for $H_X$, respectively. 
These three GPD subclasses have a one-to-one correspondence with the respective Weibull, Gumbel (exponential) and Fr\'{e}chet distributions in generalized extreme value (GEV) theory for the sample maximum $M_X={\rmmax}(X_1,\ldots, X_n)$ \cite{orju2007}, \cite{grad2010}, as defined by
\begin{align}
F^*_{M_X}(m_X) = \lim_{\theta \rightarrow \gamma} \exp \left [ - \left ( 1 + \theta \frac{m_X-\mu_{M_X}}{\sigma_{M_X}} \right )^{-\frac{1}{\theta}} \right ]
\label{eq:GEV}
\end{align} 
i.e., with the same $\gamma$ as in (\ref{eq:GPD}). The GEV framework is typically used in
block based characterization of the maximum-value distribution, while GPDs for POT make more economic use of all available data by retaining also quasi-extreme values. 

For $\chi^2_2$ i.i.d. $U_i$, the GPD of $H^{+^\prime} \stackrel{\Delta}{=}{U^+}^\prime-u^\prime_{\rmth}\geq 0$ is exponential. 
This follows from
$U^\prime \sim \exp[ - ( u^\prime_{\rmth} + {h}^{+^\prime} ) / 2] / 2$,
yielding the cumulative distribution function (CDF) of $H^{+^\prime}$ as
\begin{align}
F_{H^{+^\prime}}(h^{+^\prime}|u^\prime_{\rmth}) &= \frac{F_{U^\prime}( u^\prime_{\rmth} + h^{+^\prime}) - F_{U^\prime}(u^\prime_{\rmth})}{1-F_{U^\prime}(u^\prime_{\rmth})}\nonumber\\
&= 1 - \exp ( - {{h}^{+^\prime} }/{ 2 } )
\label{eq:exp_height_chisq_CDF}
\end{align}
i.e., (\ref{eq:GPD}) with $\gamma=0$ for $u^\prime_{\rmth}\rightarrow +\infty$. Hence
\begin{align}
f_{{H}^{+^\prime}}({h}^{+^\prime} | u^\prime_{\rmth} \rightarrow +\infty) \rightarrow \exp ( -{h}^{+^\prime}/2 ) /2
.
\label{eq:exp_height_chisq}
%\label{eq:PDF_Hpls_1}
\end{align}
The result (\ref{eq:exp_height_chisq}) can alternatively be derived based on a limit processes and the asymptotic probability law $H^{+^\prime} \stackrel{{\cal L}}{\longrightarrow} W^2_1 = \lambda^\prime_2 u^\prime_{\rmth} (L^+)^2/4$ for $u^\prime_{\rmth} \rightarrow +\infty$ \cite[Thm. 3.2]{aron1986}.

\subsubsection{Correlated Continuous Field\label{sec:excursionheight_corr}}
For a time-continuous $U(t)$ instead of discrete samples $U_i$, discrete instances of exceedance are replaced by continuous excursions of duration $\ell^+_i$. 
Within each excursion $i$, i.e., for  $t^+_i < t < t^+_i +\ell^+_i$, the excess height $H^+_i(t)$ is random but strongly correlated among nearby time points because $\ell^+_i \ll t_c$, {\em a fortiori} when $u^\prime_{\rmth} \gg 1$.
The POT can then be restricted by retaining only a single representative value $h^+_i$ at some $t_i $ per excursion,\footnote{
Note that (\ref{eq:exp_height_chisq}) applies to independent samples, viz., for $t_k \gg t_c$ spread across well separated excursions where the $K_j(t)$ in (\ref{eq:defSlepianChiSq_up}) are dominant, rather than within excursions.} e.g., the local maximum per excursion for the purpose of EMC immunity testing. 
For convenience, the midpoint $t_i \stackrel{\Delta}{=} t^+_i+ \ell^+_i/2$ may be  chosen, which is close to the instance of the local maximum of $H^+$ in the SKM when $u^\prime_{\rmth} \gg 1$ (cf. Sec. \ref{sec:excessarea}). Nevertheless, in principle any choice of $t_i \in~ ]t^+_i,t^+_i+\ell^+_i[$ is permissible, yielding (\ref{eq:exp_height_chisq_CDF}). 

\begin{figure}[!ht] 
\begin{center} \begin{tabular}{c}
\vspace{-3.7cm}
\hspace{-0.8cm}
%FIG 1 IN excessheight_upcrosss_pdf_MC_v1.m
\includegraphics[scale=0.49]{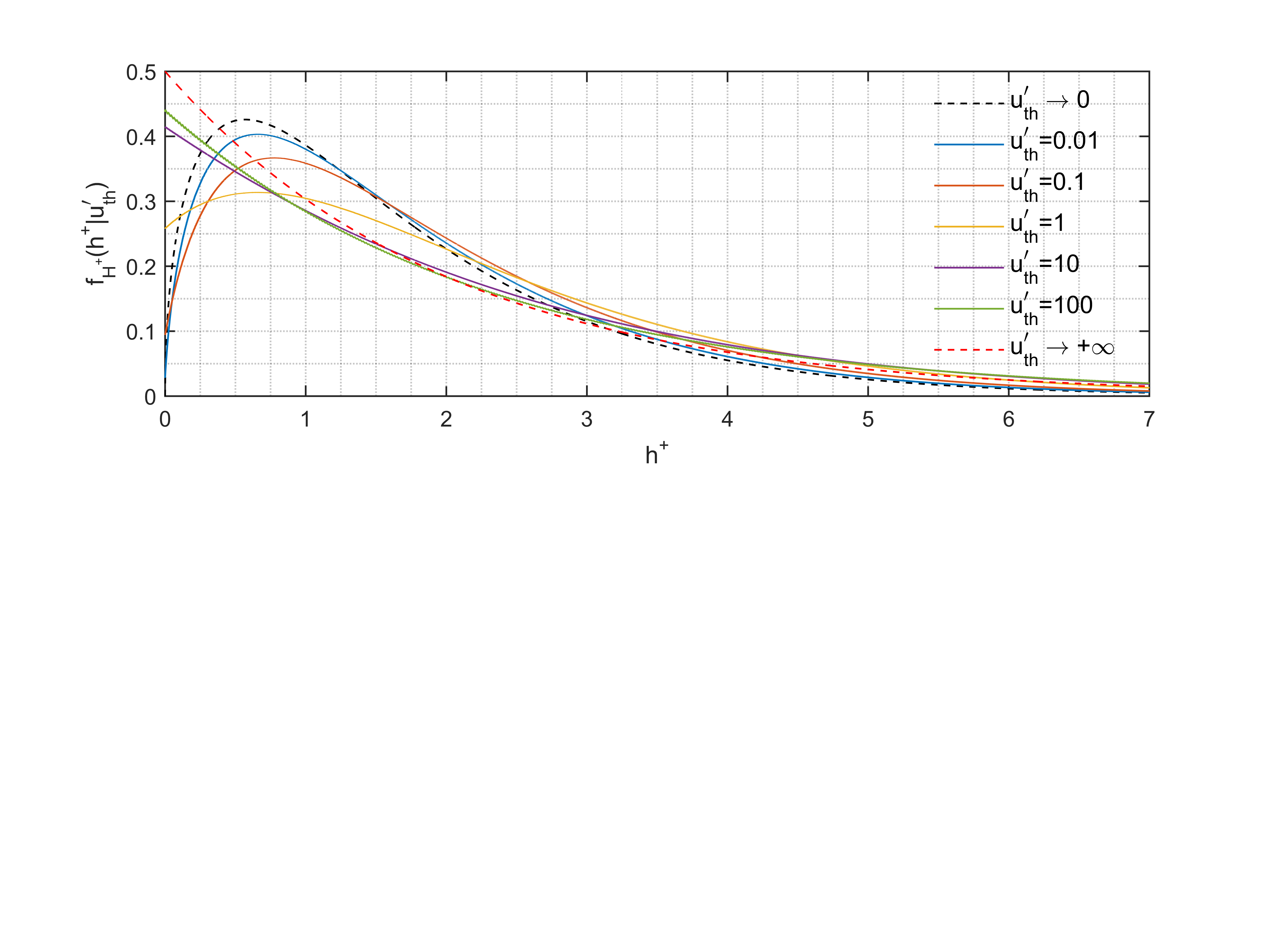}\ \\
\end{tabular} \end{center}
{\small 
\caption{\label{fig:excessheight_PDF}
\small PDF of standardized ${\tilde{H}^+}$ for selected values of $u^\prime_{\rmth}$.
}
}
\end{figure}

In order to characterize $H^+$ nonasymptotically, the general PDF of $U^+(t|u^\prime_{\rmth})$ for $t-t^+ \not \rightarrow +\infty$ can be sought from (\ref{eq:defSlepianChiSq_up}). 
In a first approximation, the asymptotic $\chi^2_2$ PDF (\ref{eq:exp_height_chisq}) can be maintained but replacing $\sigma_{H^{+^\prime}}  = \sigma_{U^\prime} = \langle U^\prime \rangle$ with $\langle H^{+^\prime}\rangle$ at $t^* \stackrel{\Delta}{=} t^+ + \langle L^+ (u^\prime_{\rmth})\rangle / 2  \ll t_c$.
 Denoting ${H^{*^\prime}} \stackrel{\Delta}{=} {H^{+^\prime}} (t^*)$, then 
\begin{align}
\langle {H^{*^\prime}} \rangle = \frac{\pi}{2} + \frac{3\pi}{2u^\prime_{\rmth}},\,\,\,\, 
\sigma^2_{H^{*^\prime}} = ({4-\pi}){\pi} + \frac{\pi^2}{u^{\prime}_{\rmth}}
\label{eq:avgHcorr}
\end{align}
for $u^\prime_{\rmth} \gg 1$ to leading order in $1/u^\prime_{\rmth}$, as follows from (\ref{eq:alphabetaU})--(\ref{eq:mean_Upls}) and (\ref{eq:avgstd_Lpls_1}). Hence $\langle H^{*^\prime}\rangle < \langle U^\prime (t-t^+\rightarrow +\infty )\rangle \equiv 2$ for $u^\prime_{\rmth} > 3\pi/(4-\pi)$.
Note that (\ref{eq:avgHcorr}) is independent of the bandwidth $\sqrt{\lambda^\prime_2}$.
As a further refinement, the PDF of the regression $\tilde{U}^+(t|u_{\rmth})$ can be used, upon setting all $K(t)=0$ in (\ref{eq:defSlepianChiSq_up}). 
For $n=2$, the PDF of the regressed height $\tilde{H}^+=\tilde{U}^+-u_{\rmth}$ follows from (\ref{eq:defSlepianChiSq_up}) after variate transformation as
\begin{align}
&~f_{\tilde{H}^+} (\tilde{h}^+|u_{\rmth}) = 
\frac{\exp\left [ - (\alpha^2 + u_{\rmth}) / (2 \beta^2)\right ]}{\sqrt{8 \pi}\, |\beta|^3} \exp\left ( - \frac{\tilde{h}^+}{2 \beta^2} \right ) \nonumber\\
&~~~~~~~~~\times
\int^{\sqrt{\tilde{h}^+ + u_{\rmth}}}_{\alpha} \frac{(y - \alpha)  \exp \left ( {\alpha}\,y / {\beta^2} \right )} {\sqrt{\tilde{h}^+ +u_{\rmth} - y^2}} \rmd y
\label{eq:PDF_Upls_firstregress}
\end{align}
for $0 \leq t - t^+ < t_c$, evaluated in $t=t^*$.
Fig. \ref{fig:excessheight_PDF} shows the numerically computed PDF (\ref{eq:PDF_Upls_firstregress}) for selected $u^\prime_{\rmth}$, indicating a transition from a $\chi^2_2$ PDF to a $\chi^2_3$ PDF when $u^\prime_{\rmth}$ decreases from $+\infty$ to $0$. Both these limit PDFs are easily retrieved analytically from (\ref{eq:PDF_Upls_firstregress}). Note that the PDF of $\tilde{H}^+|0+\equiv\tilde{U}^+|0+$ differs from the PDF of the unconditioned $U$.
%**************FIGURE 04****************
%**************FIGURE 04****************

\begin{figure}[!ht] 
\begin{center} \begin{tabular}{c}
\vspace{-3.7cm}
\hspace{-0.9cm}
%FIGURE 2006 in acf_v5.m
\includegraphics[scale=0.49]{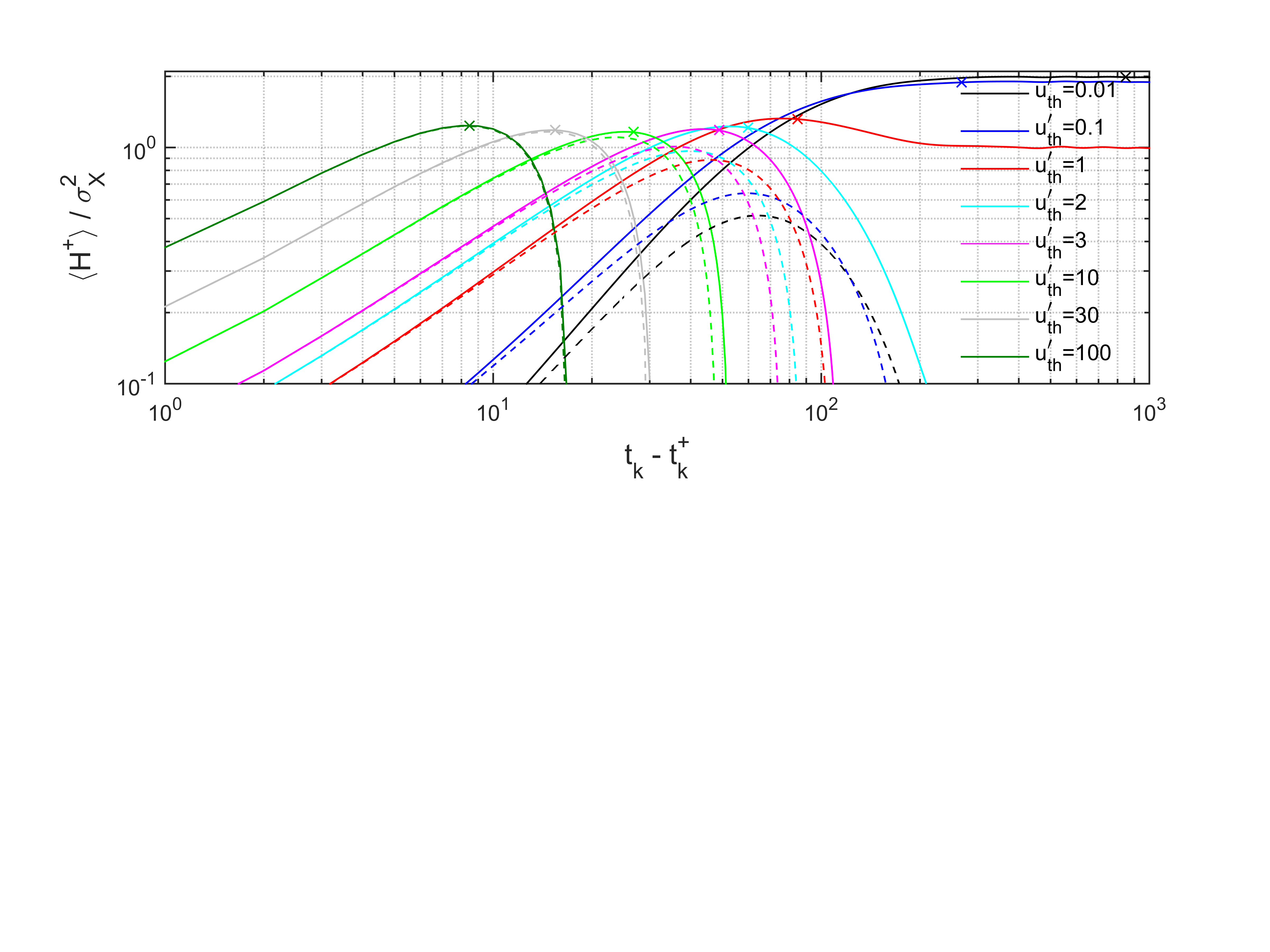}\ \\
\end{tabular} \end{center}
{\small 
\caption{\label{fig:SlepianU_avgHpls}
Normalized mean excess intensity $\langle{H}^{+^\prime}\rangle$ (solid) and its regression approximation $\langle\tilde{H}^{+^\prime}\rangle$ (dashed) for $n=2$ as a function of $t_k-t^+_k$ (in units $\Delta t_k$), after upcrossing of level $u^\prime_{\rmth}$ at $t^+_k$ and based on experimental $r^\prime_U(t_k)$. 
Cross symbols represent values at $t^*_k-t^+_k=\langle L^+ (u^\prime_{\rmth}) \rangle / 2$ situated near the expected local maxima of $H^{+^\prime}(t_k-t^+_k)$. 
}
}
\end{figure}

The normalized regression mean and variance of $\tilde{H}^{+^\prime}$ 
are
\begin{align}
\langle {\tilde{H}^{+^\prime}} \rangle = \langle {\tilde{U}^{+^\prime}} \rangle - u^\prime_{\rmth},    \,\,\,\,\,\,\,\
\sigma^2_{{\tilde{H}^{+^\prime}}} 
=
\sigma^2_{{\tilde{U}^{+^\prime}}} 
\end{align}
where $\langle {\tilde{U}^{+^\prime}} \rangle$ and $\sigma^2_{{\tilde{U}^{+^\prime}}}$ are obtained by setting $\langle K^2(t) \rangle = 0$ in (\ref{eq:mean_Upls}) and (\ref{eq:var_Upls}), respectively. 
Fig. \ref{fig:SlepianU_avgHpls} shows $\langle \tilde{H}^{+^\prime} (t-t^+) \rangle$ for the data with $n=2$. 
When $u^\prime_{\rmth} < 2$, the mean increases from zero at $t_k=t^+_k$ to the unconditioned mean $\langle U^\prime \rangle=2$ for the steady state ($t_k-t^+_k \rightarrow +\infty$) of the $\chi^2_2$ PDF, except for an offset $-u^\prime_{\rmth}$. The approximate maximum $\langle {\tilde{H}^{*^\prime}} \rangle$ decreases for increasing $u^\prime_{\rmth}$ even when $u^\prime_{\rmth} \not \gg 1$. 
For $u^\prime_{\rmth}>2$, the parabolic shape of $\langle {H^+}^\prime (t_k-t^+_k)\rangle$ (cf. Sec. \label{sec:shape}) remains approximately constant. 
Thus, for $u_{\rmth} > \langle U \rangle \equiv 2 \sigma^2_X$, long-term correlation causes $\langle H^+(t_k>t^*_k)\rangle$ to bend back toward zero, whereas for $u_{\rmth} < \langle U \rangle$ the effect of correlation on $\langle H^+\rangle$ persists as a long-term offset.
Recall that these graphs represent the time evolution and dependence of the mean $\langle {H^+}^\prime (t-t^+)\rangle$, rather than ${h^+_i}^\prime(t-t^+_i)$ for an individual excursion $i$. 
%**************FIGURE 05****************
%**************FIGURE 05****************

\subsection{Excursion Shape (Excess Intensity Profile)\label{sec:shape}}
Based on (\ref{eq:acfU}), for $u^\prime_{\rmth} \gg 1$ an excursion ${U^+}^\prime$ spanning $[ t^+,\,t^+ + L^+]$ asymptotically approximates a parabolic cap that is deterministic in curvature but random in location and height
specified by its apex
\cite{aron1986}, i.e., with upcrossing ${U}^\prime(t^+) = u^\prime_{\rmth}$ and apex $U^\prime_{\rmmax}(t_{\rmmax}) = u^\prime_{\rmth} + Z^2$ at $t_{\rmmax} = t^+ + Z/\sqrt{\lambda^\prime_2 u^\prime_{\rmth}}$. Hence the asymptotic form of ${U^+}^{\prime}(t)$ 
conditioned on its upcrossing $u^\prime_{\rmth}$ at $t^+$
is
\begin{align}
{U^+}^{\prime}(t|u^\prime_{\rmth}) \rightarrow u^\prime_{\rmth} - \left [ \sqrt{\lambda^\prime_2 \, u^\prime_{\rmth}} \, (t-t^+) - Z \right ]^2 + Z^2.
\label{eq:Slepian_up}
\end{align}
From the local maximum associated with an upcrossing-apex pair, the sample value of $Z$ follows by differentiating (\ref{eq:Slepian_up}) as
\begin{align}
z_i = \sqrt{\lambda^\prime_2 \, u^\prime_{\rmth}} \, (t_{i,{\rmmax}}-t^+_i) \simeq \sqrt{u^\prime_{i,{\rmmax}}-u^\prime_{\rmth}}
.
\label{eq:z}
\end{align}
The second equality in (\ref{eq:z}) is an approximation because, in this SKM, conditioning is based on upcrossings, as opposed to, e.g., the local maximum during the excursion \cite{lind1982}.

In general, for $d$ field randomizations ($d\geq 1$), the parabolic excursion generalizes to a paraboloid cap with a $(d+1)$-D (hyper)volume of excess above a $d$-D (hyper)plane threshold. 

\subsection{Excursion Area (Excess Energy)\label{sec:excessarea}}
Integration of the instantaneous excursion height across the excursion length 
$L^+=t^- -t^+$ 
yields the random excursion area $A^+ = \int^{t^-}_{t^+} h^+(t) \rmd t$.
In dynamic EMEs, the rate of the induced field fluctuations $\sqrt{\lambda^\prime_2}$ varies (e.g., stir speed in MSRCs \cite[sec. VI-B2]{arnaexcur}), thus affecting the profile of $U(t)$. Therefore, the dependence of $A^+$ on $\lambda^\prime_2$ is of specific interest.

For $u^\prime_{\rmth} \rightarrow + \infty$, in the deterministic second-order, i.e., parabolic limit SKM, the product of the duration (chord length) and the maximum height multiplied by $2/3$ provides an estimate of $A^+$. 
The total area $A$ below $U(t)$ spanning $L^+$ (cf. Fig. \ref{fig:diagram_Apls}) is then the sum of $A_{\rmth}$ for the rectangle of fixed height $u_{\rmth}$ and random width $L^+$ plus $A^+$ for the parabolic cap of random height $H$ and width $L^+$, 
i.e., 
\begin{align}
A &= A_{\rmth} + A^+ = u_{\rmth} L^+ + {2 H^+ L^+ }/{3} 
.
\end{align}
For $H^{+^\prime}$, the asymptotic PDF (\ref{eq:exp_height_chisq}) can be used.
With 
${A^+}^\prime\stackrel{\Delta}{=} A^+/\sigma^2_X \stackrel{{\cal L}}{\rightarrow} {4 W^3_1}/(3\sqrt{\lambda^\prime_2 \, u^\prime_{\rmth}})$ 
and
\cite[eq. (3.326.2)]{grad1}, the asymptotic PDF of $A^{+^\prime}$ follows by variate transformation as
\begin{align}
&\, f_{A^{+^\prime}}(a^{+^\prime}) \rightarrow \left ( \frac{\lambda^\prime_2 \, u^\prime_{\rmth}}{48\, a^{+^\prime}} \right )^{\frac{1}{3}} \exp \left [ - \left ( \frac{9 \lambda^\prime_2 \, u^\prime_{\rmth}}{128} ( a^{+^\prime} )^2 \right )^{\frac{1}{3}}  \right ] \nonumber
\\
&=
\left ( \frac{\pi}{6 \langle A^{+^\prime}\rangle^2 a^{+^\prime}} \right )^{\frac{1}{3}} \exp\left [- \left ( \frac{9 {\pi} ( a^{+^\prime}  )^2}{16\langle A^{+^\prime}\rangle^2} \right )^{\frac{1}{3}}\right ]\label{eq:PDF_Apls}
\end{align}
with the asymptotic mean and standard deviation of $A^{+^\prime}$ as
\begin{align}
\langle A^{+^\prime} \rangle = \sqrt{\frac{8\pi}{\lambda^\prime_2 \,  u^\prime_{\rmth}}},~~~
\sigma_{A^{+^\prime}} = \sqrt{\left ( \frac{256}{3} - 8\pi\right ) \frac{1}{\lambda^\prime_2 \, u^\prime_{\rmth}}}
\label{eq:PDF_musigmaA}
.
\end{align}
The coefficient of variation
$\nu_{A^+} 
= \sigma_{A^+} / \langle A^+ \rangle 
= \sqrt{32/(3\pi)-1}\simeq 1.55 > 1$ indicates a considerably larger relative spread of fluctuations of $A^+$ than those of $\tilde{H}^+$, viz., $\nu_{\tilde{H}^+}=\sqrt{2/q} < 1$ for $2 < q < 3$ depending on $u^\prime_{\rmth}$.

\begin{figure}[!ht] 
\begin{center} \begin{tabular}{c}
\vspace{-3.7cm}
\hspace{-0.8cm}
%FIG 2 in excessarea_Apls_pdf_theo_v2.m
%\includegraphics[scale=0.67]{Fig06_fA_hplsmax1e4_resH1e2_resY1e3}
\includegraphics[scale=0.67]{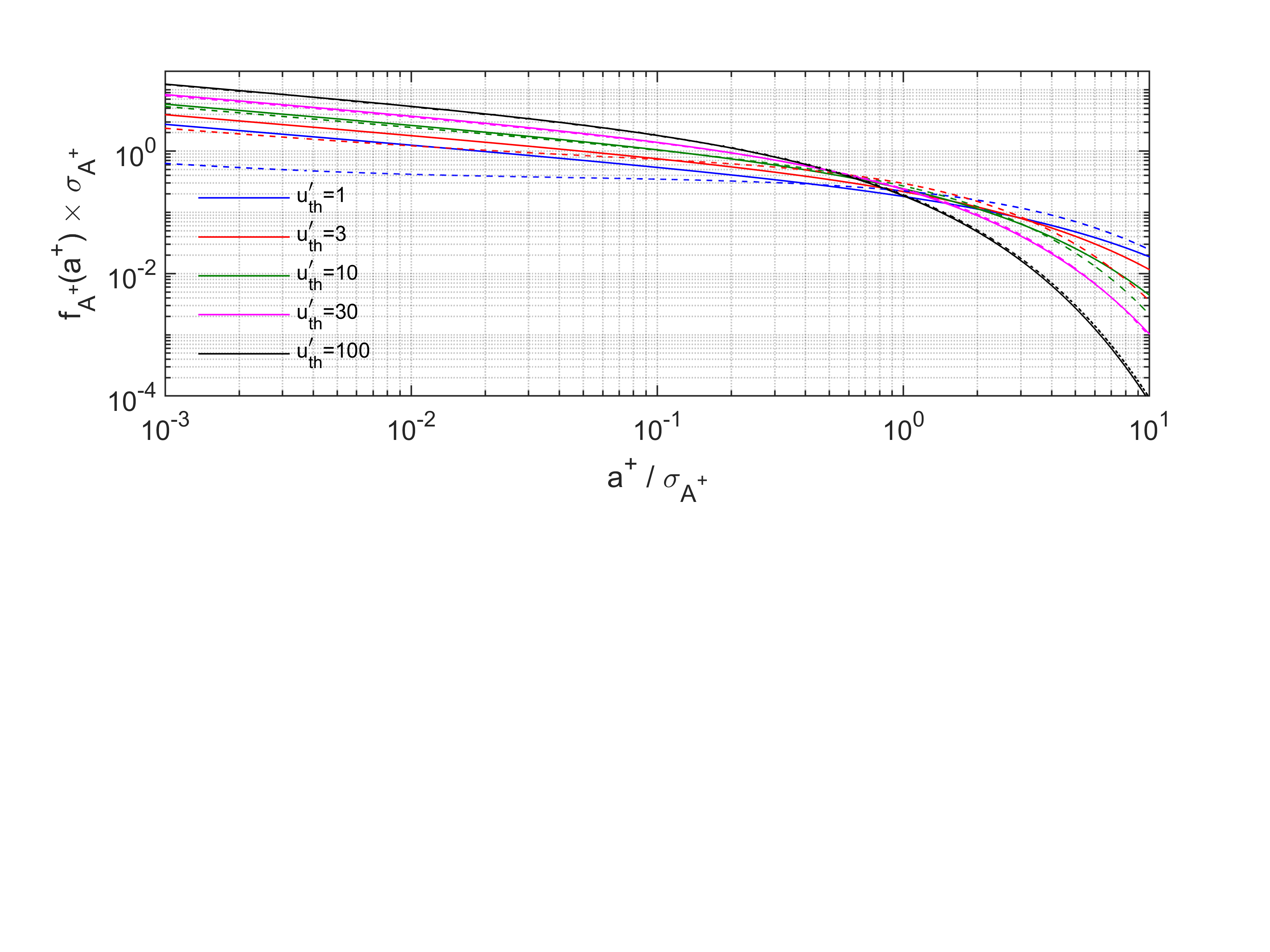}
\end{tabular} \end{center}
{\small 
\caption{\label{fig:PDF_Apls}
\small
Asymptotic $\chi^3_2$ PDFs of $A^+$ (\ref{eq:PDF_Apls}) (solid) and regression PDFs  of $\tilde{A}^+$ (\ref{eq:PDF_Apls_firstregress}) (dashed)
for selected values of $u^\prime_{\rmth}$.
}
}
\end{figure}

Eq. (\ref{eq:PDF_Apls}) represents a $\chi^3_2$ (chi-cubed with two degrees of freedom) 
asymptotic PDF for the scaled product of a $\chi_2$-distributed $L^+$ and a fully dependent $\chi^2_2$-distributed $H^+$. 
Fig. \ref{fig:PDF_Apls} shows the scaled (\ref{eq:PDF_Apls}) for selected $u^\prime_{\rmth}$.
Because of the asymptotic Gaussianity of $U$ above $u^\prime_{\rmth} \gg 1$ \cite{arnalocavg}, (\ref{eq:PDF_Apls}) also holds (with a scaled mean and variance) for the excursion area of the Gaussian $X(t)$ itself and for more general $\chi^2_n$ intensities, including $\chi^2_1$ for the intensity of random static fields. 
%**************FIGURE 06****************
%**************FIGURE 06****************

For high but non-asymptotic $u^\prime_{\rmth} \gg 1$, excursions remain sparse and short whence the terms $K_j(t)$ dominate the sum in (\ref{eq:defSlepianChiSq_up}). For $n=2$, the asymptotic $\chi^2_2$ PDF of $H^+$ tends increasingly toward $\chi^2_3$ when $u^\prime_{\rmth}$ is lowered (cf. Sec. \ref{sec:excursionheight_corr}). 
The corresponding perturbation and evolution of $A^+$ depends on $r_U(t)$ -- which affects the correlation between the $H^+$ of excursions and between their $L^+$ \cite{sire2007} -- and on the decreasing correlation between $H^+$ and $L^+$ (reduced curvature). 
In a first approximation, the parabolic approximation $A^+ = 2 H^+ L^+/3$ can be maintained on account of a similar shape of $\langle H^+ \rangle$ for $u^\prime_{\rmth} > 2$ (cf. Fig. \ref{fig:SlepianU_avgHpls}), with $\sigma_{U^+} \ll \langle U^+ \rangle$ and $\rho_U(t-t^+\ll t_c) \simeq 1$ when $u^\prime_{\rmth} \gg 1$ (cf. Fig. \ref{fig:Slepian_musigma}). Thus, {\em individual\/} excursions are also approximately parabolic 
within each excursion $[t^+_i,\/ t^+_i + \ell^+_i]$, while $H^+$ and $L^+$ are still strongly correlated.
Application of (\ref{eq:avgHcorr}) then yields 
\begin{align}
\langle A^{*^\prime} \rangle \stackrel{\Delta}{=} \left \langle \frac{2 H^{*^\prime} L^+ }{ 3 }  \right \rangle &\simeq 
\sqrt{\frac{\pi^3}{2\lambda^\prime_2 u^\prime_{\rmth}}
}
\label{eq:avgAstar}
\end{align} 
%\end{document}
as a nonasymptotic estimate\footnote{Note that $\langle L^+ (u^\prime_{\rmth} \not \gg 1) \rangle > \langle L^+ (u^\prime_{\rmth} \rightarrow +\infty) \rangle$, which counters the diminishing effect of $\langle H^* \rangle < \langle U \rangle$ on $A^+$.} replacing $\pi \langle A^{+^\prime} \rangle / 4$ in (\ref{eq:PDF_musigmaA}).
As a further refinement, application of the variate transformation $A^+ \stackrel{\cal L}{\rightarrow} 2 (H^+)^{3/2}/3$ to (\ref{eq:PDF_Upls_firstregress}) results in
\begin{align}
&~ f_{\tilde{A}^+} (\tilde{a}^+|u_{\rmth}) \simeq 
\frac{C_{\tilde{A}^+}}{(\tilde{a}^+)^{{1}/{3}}}
 \exp\left ( - \frac{h^+(\tilde{a}^+)}{2 \beta^2} \right ) \nonumber\\
&~~~\times
\int^{\sqrt{h^+(\tilde{a}^+) + u_{\rmth}}}_{\alpha} \frac{(y - \alpha) \exp \left ( {\alpha\, y}/{\beta^2} \right )} {\sqrt{h^+(\tilde{a}^+) + u_{\rmth} - y^2}} \rmd y
\label{eq:PDF_Apls_firstregress}
\end{align}
where $h^+(\tilde{a}^+) = (3\tilde{a}^+\sqrt{\lambda^\prime_2 u^\prime_{\rmth}}/4)^{2/3}$, with
\begin{align}
C_{\tilde{A}^+} = \frac{(\lambda^\prime_2 u^\prime_{\rmth})^{1/3} \, \exp [ - (\alpha^2 + u_{\rmth})/(2 \beta^2) ] }{{2^{11/6} \, 3^{1/3} \, \sqrt{\pi}\, |\beta|^3}}
.
\end{align} 
Combining (\ref{eq:avgAstar}) and (\ref{eq:PDF_Apls_firstregress}) yields $f_{\tilde{A}^*}(\tilde{a}^*) \rightarrow (4/\pi) f_{\tilde{A}^+}(4\tilde{a}^*/\pi)$ when $u^\prime_{\rmth} \rightarrow +\infty$.
Fig. \ref{fig:PDF_Apls} shows close correspondence between the regression PDF (\ref{eq:PDF_Apls_firstregress}) and the asymptotic PDF (\ref{eq:PDF_Apls}) when $u^\prime_{\rmth} \gg 1$, {\it a fortiori} for the pertinent range $a^+/\sigma_{A^+}\ll 1$.

The approximation in (\ref{eq:PDF_Apls_firstregress}) reflects the departure from ideal full statistical dependence between $L^+$ and $H^+$ in writing $A^+ \simeq 2 (H^+)^{3/2}/3$.
When $u^\prime_{\rmth} \not \rightarrow +\infty$, the actual dependence can be represented by a copula density linking $f_{H^+}(h^+)$ and $f_{L^+}(\ell^+)$ \cite{copula2016}. The departure of nonasymptotic $f_{L^+}(\ell^+)$ from a $\chi_2$ PDF can be evaluated numerically using a high-order Rice series for multiple crossings of $u^\prime_{\rmth}$ \cite[sec. 3.4]{rice1945}, \cite{azai2009}.

\section{Experimental Results\label{sec:exp}}
The following results are based on $40\,000$ data points measured for uniform continuous stirring in a MSRC as a dynamic EME  \cite{arnajitt}, \cite{arnaexcur}. 
A pulse amplitude modulated (PAM) source with carrier frequency $f_c = 2.5$ GHz and 40\% duty cycle provides the excitation. 
Each $k^{\rmth}$ post-processed data point $X(t_k)$ is a 998-point sample average over the late steady-state ON part received from the PAM input.
The sample times are $t^{(j)}_k \in [1000\,\delta{t},1997\,\delta{t}]$ within each $k^{\rmth}$ pulse period $k\cdot[0,T]=k\cdot[0,5000\,\delta{t}]$, with spacing $\Delta t_k = T = 100\,\mu$s.
Data were collected across one full revolution at a constant rotational speed of 0.25 r/s, for which $\sqrt{\lambda^\prime_2}=147.9$ rad/s. 

\begin{figure}[H] 
\begin{center} \begin{tabular}{c}
\vspace{-3.7cm}
\hspace{-0.7cm}
%FIG 1000 IN UCTF_v4_2D_Tektronix_NoTrfMin_FORPAPER.m
\includegraphics[scale=0.48]{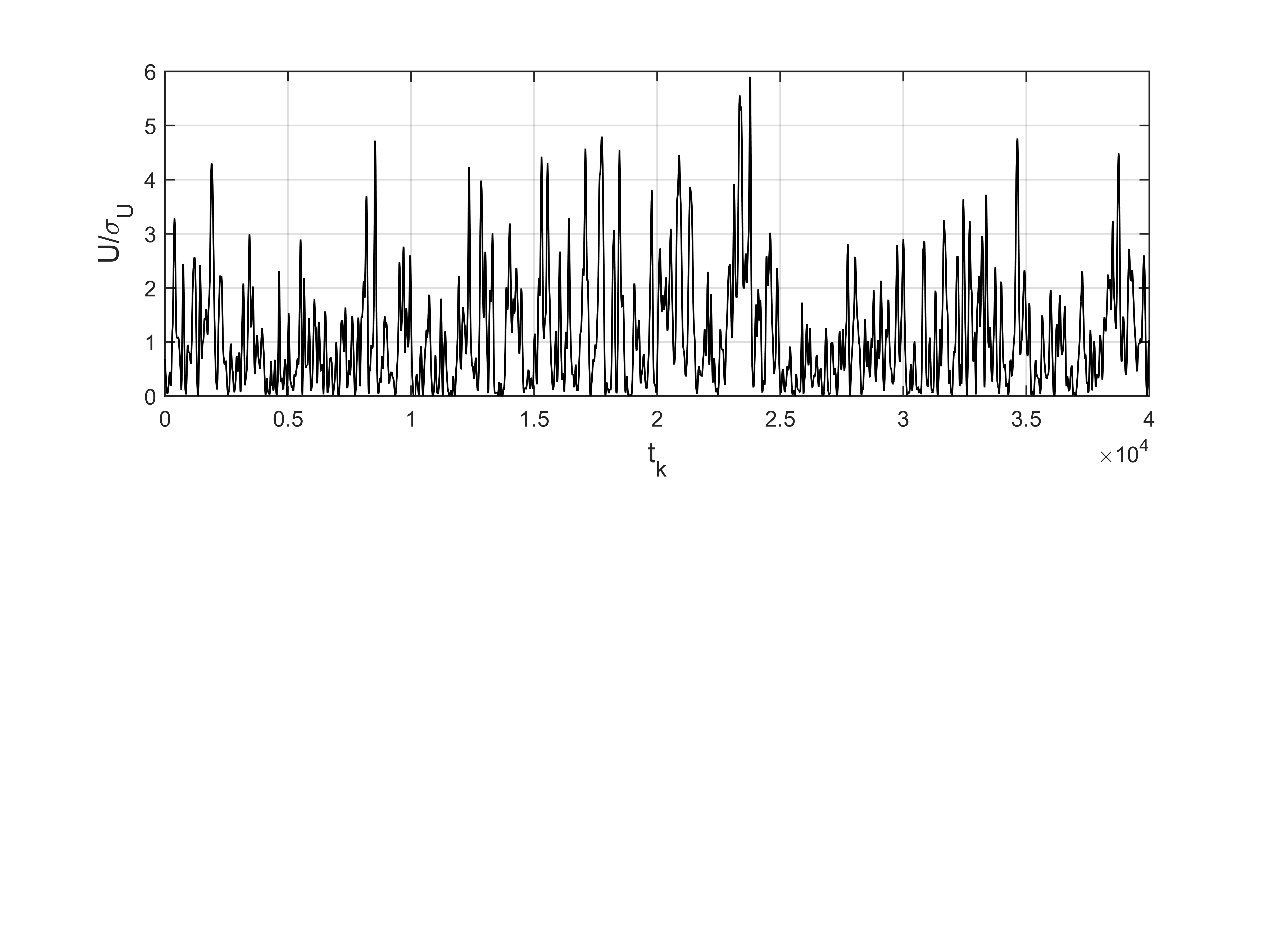}\ \\
(a)\\
\vspace{-0.2cm}
\hspace{-0.8cm}
%FIG 25010 IN Slepian_v3_Tektronix_NoTrfMin_freqstirspeed.m
\includegraphics[scale=0.48]{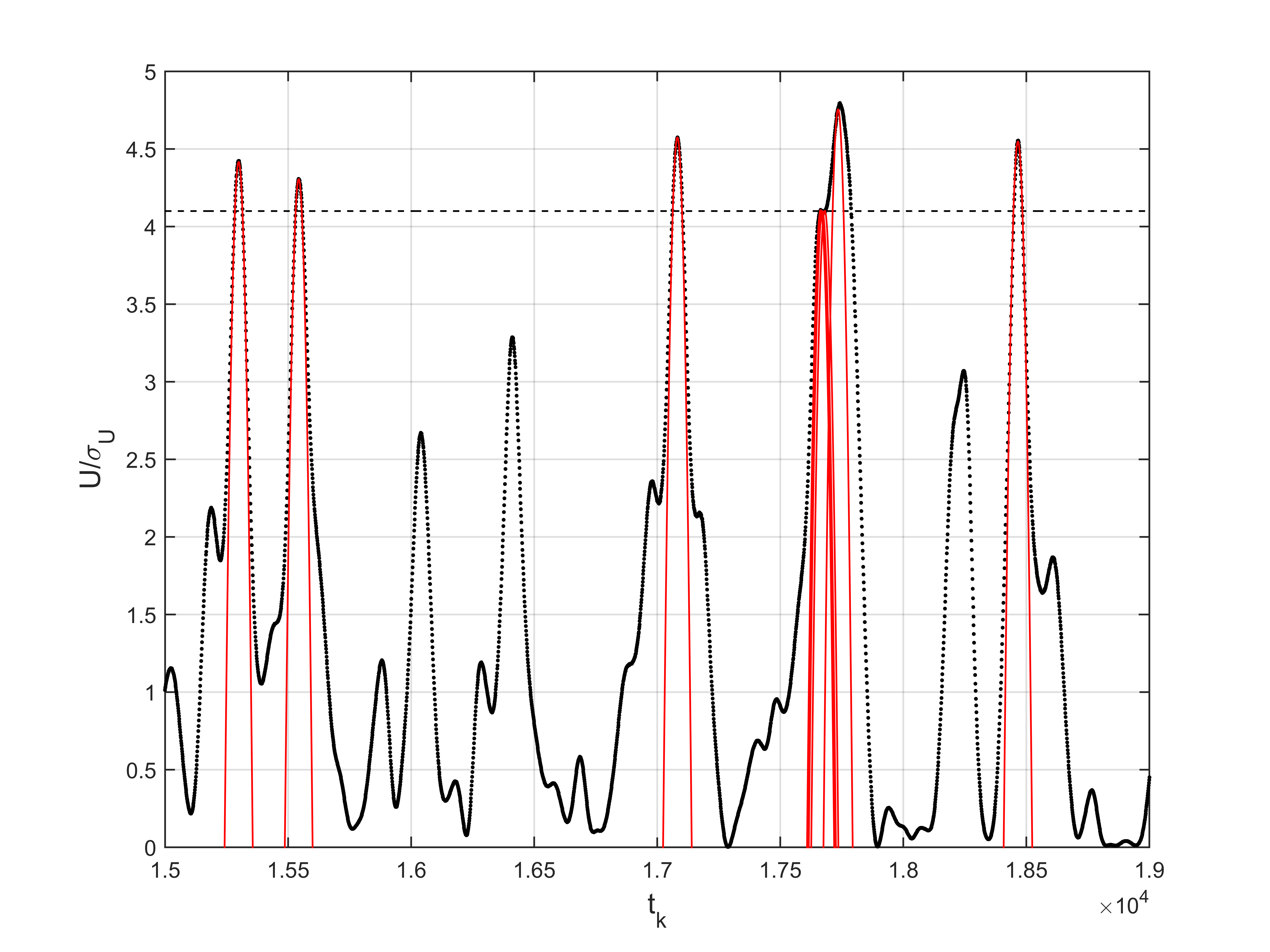}\ \\
(b)\\
\vspace{-0.2cm}
\hspace{-0.7cm}
\includegraphics[scale=0.48]{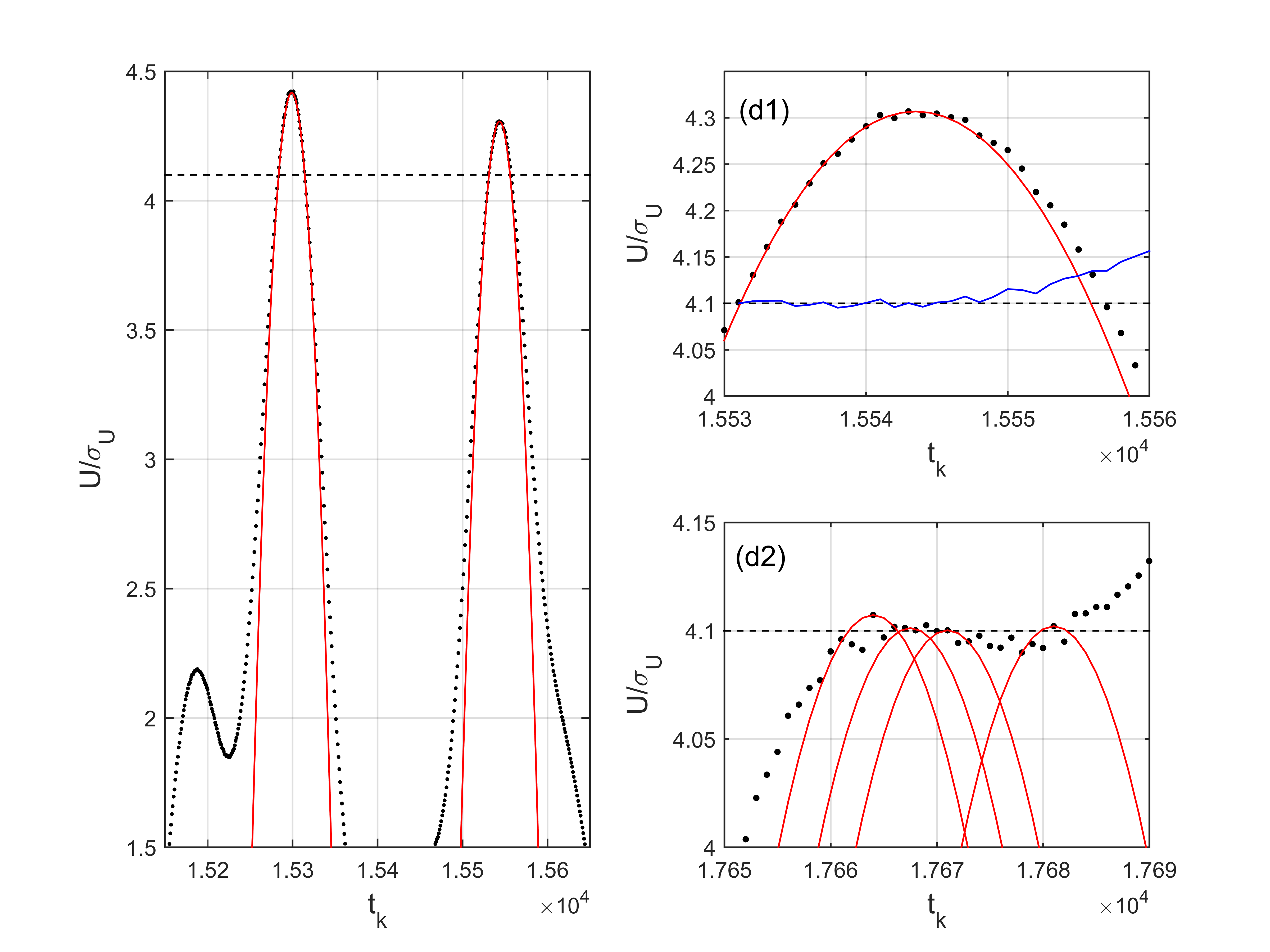}\ \\
(c)~~~~~~~~~~~~~~~~~~~~~~~~~~~~~~(d)\\
\end{tabular} 
\end{center}
{\small 
\caption{\label{fig:Slepian_Tektronix_parabolae}
\small
(a) Experimental full stir sweep data $U(t_k)/\sigma_U$.
(b)--(d) Comparison of experimental data $U(t_k)/\sigma_U$ (black dots) vs. SKM parabolic model (red curves) conditioned on upcrossings $t^+_i$ of $u_{\rmth}/\sigma_U = 4.1$: 
(c) for $15\,150 \leq t_k \leq 15\,650$, with 
$(t^+_{3},t_{3,\rmmax},z_3)=(15\,284, 15\,299, 0.449)$ and 
$(t^+_{4},t_{4,\rmmax},z_4)=(15\,531, 15\,543.5, 0.374)$; 
(d1) data (black dots), SKM conditioned on $t^+_{4}$ (red solid), and residual difference raised by $u_{\rmth}/\sigma_U$ (blue solid);
(d2) SKMs conditional on local maximum for noisy crossings of $u_{\rmth}/\sigma_U$
for 
$15\,650 \leq t_k \leq 17\,690$, at 
$(t^+_{i},t_{i,\rmmax},z_i)=
(17\,664,17\,664,0)$,
$(17\,666,17\,667.5,0.045)$,
$(17\,671,17\,671,0)$ and
$(17\,681,17\,681,0)$ with $i=6,7,8$ and $9$, respectively. All $t_k$ in units $\Delta t_k$.
}
}
\end{figure}

Fig. \ref{fig:Slepian_Tektronix_parabolae} shows $U(t_k)/\sigma_U \equiv U^\prime(t_k)/2$ and its SKM parabolic approximations above a level $u_\rmth/\sigma_U \equiv u^\prime_{\rmth}/2 = 4.1$. 
Fig. \ref{fig:Slepian_Tektronix_parabolae}(b)--(d) indicates that the SKM is accurate for this threshold provided that the local maxima are sufficiently widely separated in time.
Although the SKM is strictly valid only in excess of $u_{\rmth}$, the model appears to be adequate also well below $u_{\rmth}$.
The differences between the data and the SKM estimates, shown in blue in Fig. \ref{fig:Slepian_Tektronix_parabolae}(d1), increase on average with $t_k-t^+_k$ and demonstrate the nonstationarity of $K(t_k-t^+_k)$. At the downcrossing $t^-_k$, the error is $0.034$, i.e., $0.8\%$ of $u_\rmth/\sigma_U$.
%**************FIGURE 07****************
%**************FIGURE 07****************
The small bias of the SKM parabola towards lower $t_k$ is a result of conditioning on upcrossings as opposed to, e.g., local maxima \cite{lind1982}, \cite{blach1988}. 
A slightly improved approximation can be obtained using a more elaborate doubly conditioned SKM that uses both up- and downcrossings \cite[eq. (20)]{slep1963}.

Fig. \ref{fig:Slepian_Tektronix_parabolae}(d2) shows that noise in data -- even after averaging within each pulse period -- affects the accuracy of the SKM, {\it a fortiori} near a local maximum. 
Here, the SKM was conditioned on the estimated height and location of such a maximum. Such conditioning is prone to larger uncertainty and requires precise knowledge of $\lambda^\prime_4$ in (\ref{eq:acfU}) \cite[Figs. 6(c) and 7(c)]{arnaexcur}, compared with conditioning on upcrossings.
Clearly, the overlapping SKM parabolas fail to provide an accurate approximation. This is seen in Fig. \ref{fig:Slepian_Tektronix_parabolae}(b) and (d2) from the overlap of adjacent parabolas, each one being characterized by $z_i\simeq 0$ on account of $u^\prime_{{\rmth},i} \simeq u^\prime_{{\rmmax},i}$, whence successive crossings now resemble a noise process themselves.

The approximation of the next local maximum at $t_{{\rmmax},10}$ [cf. Fig. \ref{fig:Slepian_Tektronix_parabolae}(b)] is also affected, because at $t^+_{10}$ the data near $u^\prime_{\rmth}$ lack the smoothness of threshold crossing required for an accurate SKM.
Comparison of Fig. \ref{fig:Slepian_Tektronix_parabolae}(d1) with Fig. \ref{fig:acfXUK}(b) confirms that, for this excursion at this $u_{\rmth}$,  the excursion length $\ell^+ = 25\,\Delta t_k \ll (103.5/2)\, \Delta t_k = t_c$ is sufficiently short compared to $t_c$ in order that the regression term dominates the $K_j(t)$ in (\ref{eq:defSlepianChiSq_up}).
 
Fig. \ref{fig:Slepian_Tektronix_Zscores}(a) shows sample values of $Z$ from the SKM for excursions $i$ in Fig. \ref{fig:Slepian_Tektronix_parabolae}(a) at selected threshold levels.
When local maxima of $U(t_k)$ are closely spaced, the $z_i$ become more volatile, leading to peaks of $z_i$, e.g., for $t_k \simeq 17\,000 \, \Delta t_k$ to $23\,000 \, \Delta t_k$.
Except for such peaks, most values of $z_i$ are notably similar for a chosen $u_{\rmth}/\sigma_U$. 
%This suggests that $\langle Z \rangle$ serves as a mean scale of excursion size per threshold level. 

For an arbitrary excursion $i$, the value of $z_i$ decreases for increasing $u_{\rmth}/\sigma_U$.
Further analyzing this decrease, Fig. \ref{fig:Slepian_Tektronix_Zscores}(b) demonstrates that the functional dependence $z_i(u_{\rmth}/\sigma_U)$ exhibits the typical square-root roll-off to zero (rounded tip) as the local maximum is approached sufficiently closely, i.e., for $t_{i,{\rmmax}} -t^+_i \rightarrow 0$. Sharp notches in $z_i(u_{\rmth}/\sigma_U)$ indicate excursions that degenerate to multiple peaks as $u_{\rmth}$ is increased, e.g., near $u_{\rmth}/\sigma_U=4.1$ for $t_{7} = 17\,670\, \Delta t_k$. Such degeneracies typically start at relatively large $z_i$ ({long excursion) for low $u_{\rmth}/\sigma_U$. 
A large cluster of local maxima breaks up into smaller excursions when $u_{\rmth}$ is increased. Such clustering may have important consequences in an EM test device with a relatively large time constant or small recovery times. 
In this respect, the notches in $z_i(u_{\rmth}/\sigma_U)$ are indications of changes in connectivity (topology) of the excursions.

Fig. \ref{fig:FigP500_TektronixSV1e4CW_excurarea} shows experimental sample values of $m_{A^+}$ and $s_{A^+}$ for $\langle A^+ \rangle$ and $\sigma_{A^+}$ as a function of $u_{\rmth}/\sigma_U$. 
Each value of $a^+_i$ was estimated from a discrete staircase approximation of $U_i(t)$ by rectangles of height $U^+(t_k)$ and width $\Delta t_k$. 
For $u_{\rmth}/\sigma_U \gg 1$, the estimated values of $a^+_i$ ultimately suffer from inevitable depletion of data and extreme narrowness of excursions, whereby the accuracy of the staircase  approximation suffers.
%**************FIGURE 08****************
\begin{figure}[!ht] 
\begin{center} \begin{tabular}{c}
\vspace{-4.3cm}
\hspace{-0.9cm}
%FIGURE 2512 IN Slepian_v3_Tektronix_NoTrfMin_freqstirspeed.m
\includegraphics[scale=0.48]{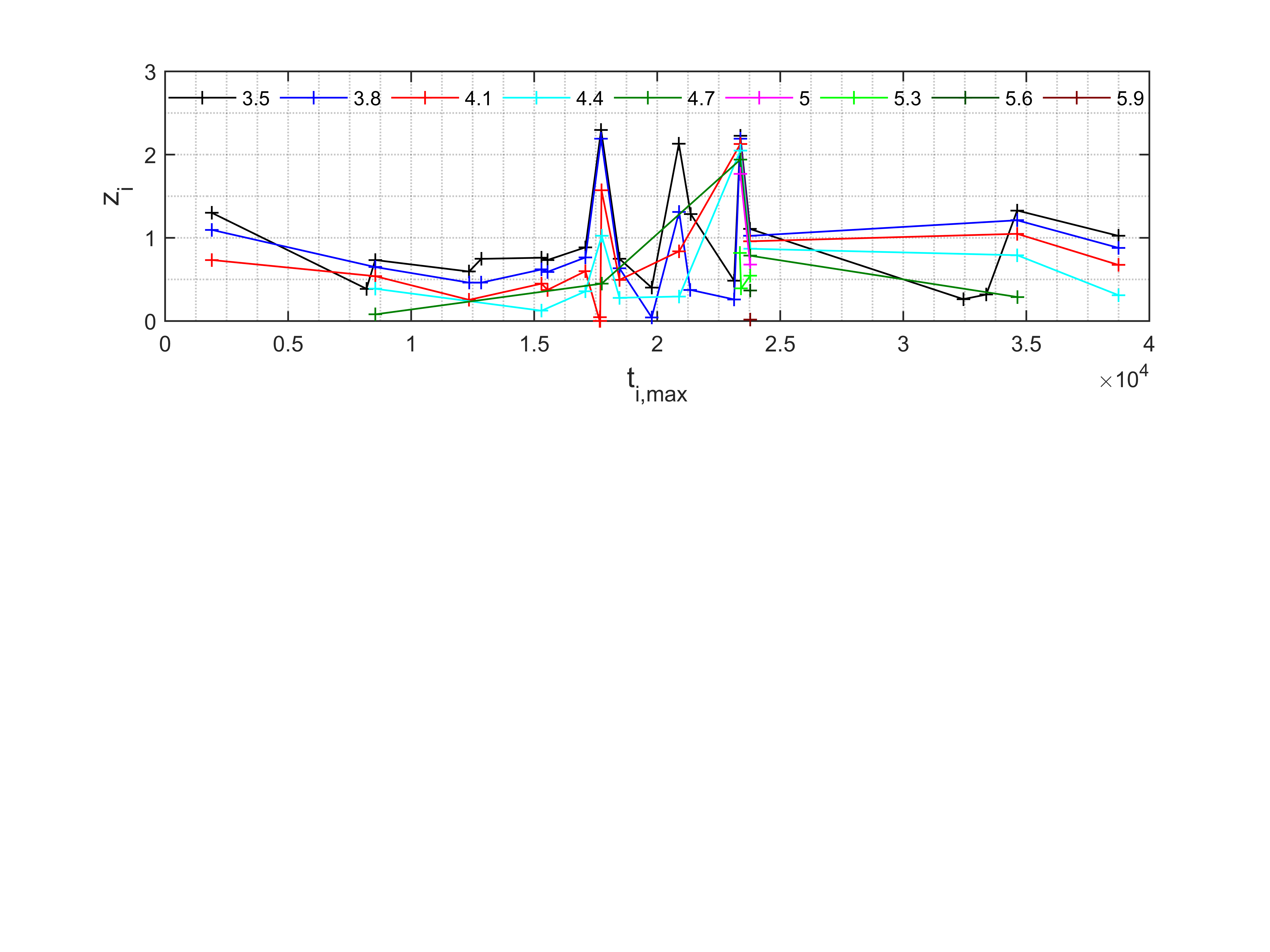}\ \\
(a)\\
\vspace{-3.7cm}
\hspace{-0.9cm}
%FIGURE 3511 IN Slepian_v3_Tektronix_NoTrfMin_freqstirspeed.m
\includegraphics[scale=0.48]{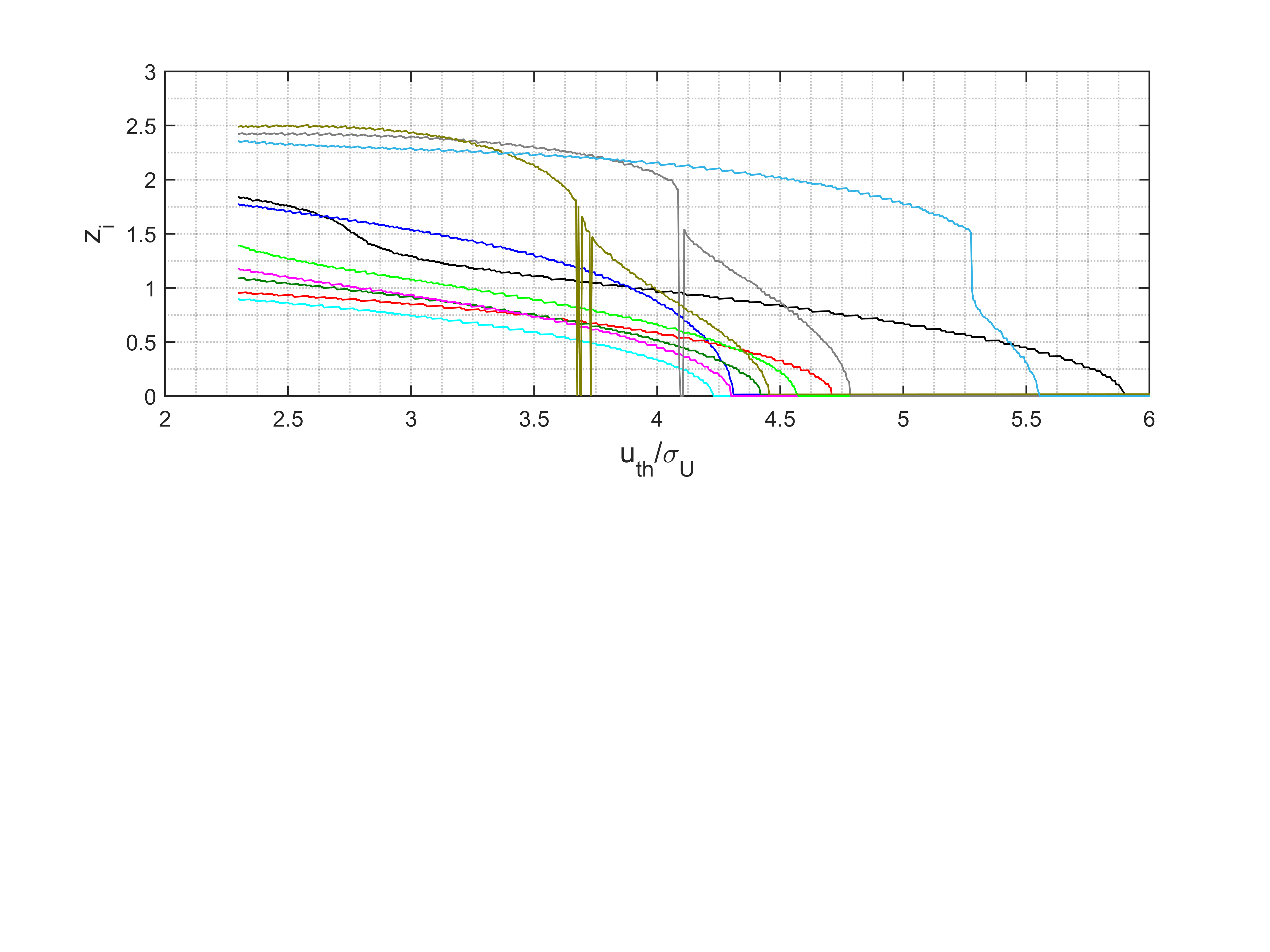}\ \\
(b)\\
\end{tabular} \end{center}
{\small 
\caption{\label{fig:Slepian_Tektronix_Zscores}
\small
Sample values $z_i$ of $Z$ in the SKM parabolic approximation of $U(t)/\sigma_U$:
(a) for $i^{\rmth}$ positive excursion at selected threshold levels $u_{\rmth}/\sigma_U$ from 3.5 to 5.9;
(b) sample paths $z_i(u_{\rmth}/\sigma_U)$ for selected excursions in Fig. \ref{fig:Slepian_Tektronix_parabolae}(a).
}
}
\end{figure}
%**************FIGURE 08****************
%**************FIGURE 09****************
\begin{figure}[!ht] 
\begin{center} \begin{tabular}{c}
\vspace{-3.7cm}
\hspace{-0.9cm}
%FIGURE 501 IN UCTF_v4_2D_Tektronix_NoTrfMin_FORPAPER.m
\includegraphics[scale=0.48]{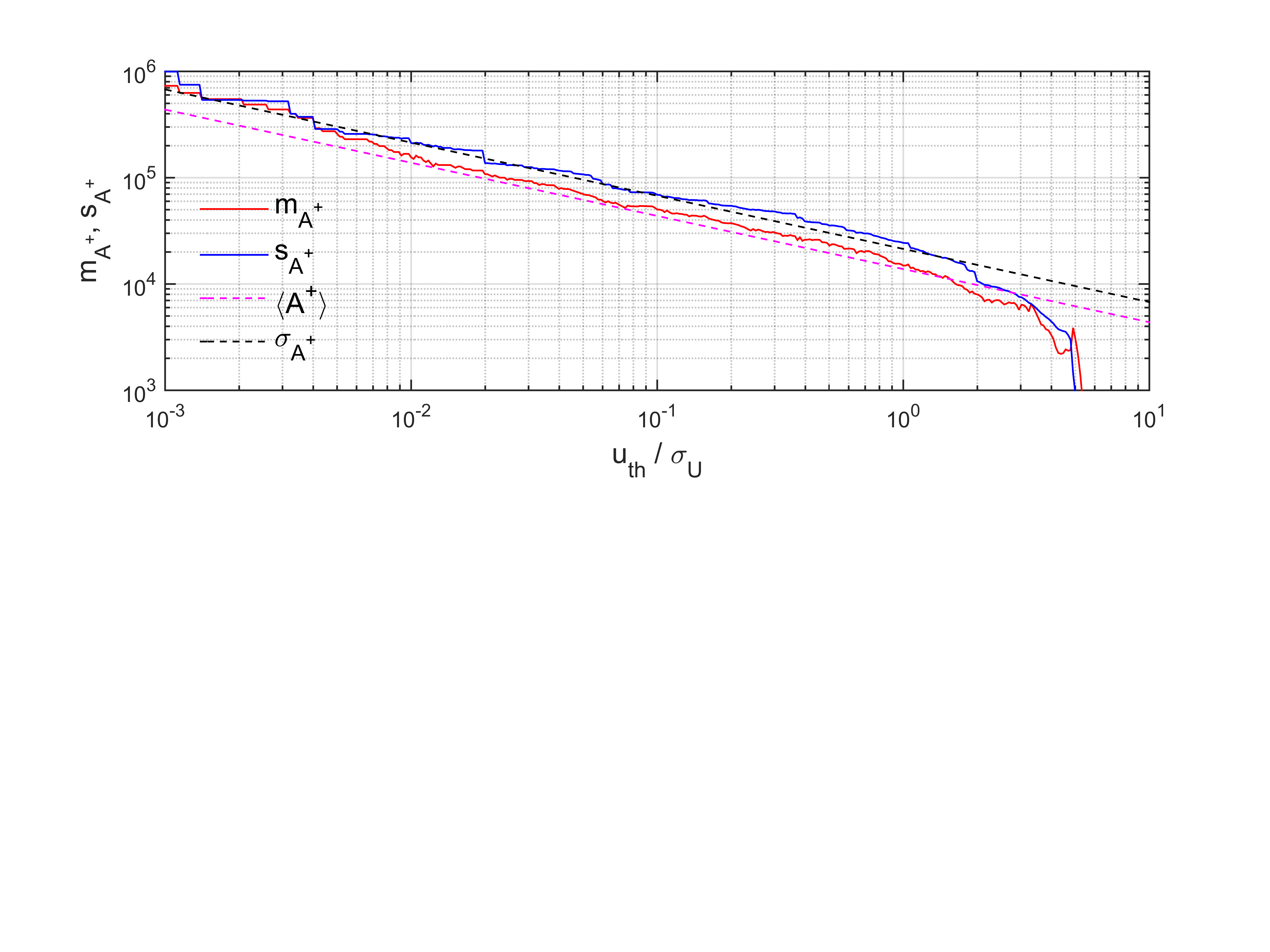}\ \\
\end{tabular} \end{center}
{\small 
\caption{\label{fig:FigP500_TektronixSV1e4CW_excurarea}
\small Experimental mean $m_{A^+}$ and standard deviation $s_{A^+}$ (solid red resp. blue) in dimensionless units $1^2 \cdot \Delta{t_k}$  vs. theoretical asymptotic $\langle{A^+}\rangle$ and $\sigma_{A^+}$ (dashed magenta resp. black) as a function of $u_{\rmth}/\sigma_U$. 
}
}
\end{figure}
%**************FIGURE 09****************

\section{Excess Energy and Power in Dynamic EMEs\label{eq:dynamicpower}}
\subsection{Hybrid Modulation}
In this second part of the paper, the interest is shifted to the characterization of energy and power exceedance in a rapid dynamic, i.e., continuously varying EME. 
Its variation can occur with respect to the electronic state (e.g., transitions of ambient EM sources or biasing fields) or the mechanical configuration (e.g., motion of EM cavity boundaries or scatterers).
Configurational changes induce secondary modulation in amplitude \cite{arna2007}, phase or frequency \cite{arnajitt} to an unmodulated continuous wave (CW) or to a primary (i.e., electric heterodyne) modulated source signal field. The occurrence of both modulation types produces {\it hybrid modulation} (HM).
The overall rate of fluctuation of the field in HM is governed by both the source and the time-varying EME \cite{arnamaxratefluct2005}, \cite{arna2007b}. 

In focusing on dynamic effects, the source is here assumed to be unmodulated (CW), including the limiting case of a narrowband modulation, while the EME provides additional harmonic variation at a constant rate. In a MSRC context, overlap of cavity modes causes a CW source field to become narrowband modulated, while mode stirring produces modulation of each modal component. 
Thus, the spectra reduce to a single line in each case. The HM can be characterized by the HM index
\begin{align}
m\stackrel{\Delta}{=}\sqrt{\lambda^\prime_2}/\omega
\label{eq:HMindex}
\end{align}
representing the ratio of the EME's dynamic (stir) rate of fluctuation $\sqrt{\lambda^\prime_2}$ to the rate of the static (unstirred) EM oscillation $\omega$ of the CW excitation source.

For wideband electric modulation of the source signal, HM applies to each quasi-static frequency component of the source and produces a widened dynamic spectrum. HM by the EME results in an increased bandwidth. 
Similarly, for an EME changing at a nonuniform rate \cite{arnajitt}, its effect on the EM field can be characterized by the configurational Fourier spectrum, in which the $i$th spectral component of the EME transfer function can be analyzed in conjunction with the $j$th component of the EM source.
If $\sqrt{\lambda^\prime_2} \sim \omega$ or if either spectrum is sufficiently wideband such that $\sqrt{(\lambda^\prime_2)_i} \sim \omega_j$, then HM coupling occurs. 

\subsection{Effective Power}
The previous exposition for quasi-static operation can be extended to dynamic EMEs.
Recall the circuit definitions of deterministic instantaneous power $p(t)$, energy $u(T)$ and quasi-static effective (time-averaged) power $p_{\rmeff}$, i.e.,
\begin{align}
p(t) = v(t) i(t),~~
u(T) = \int^T_0 p(t) \rmd t,~~
p_{\rmeff} = \frac{u(T)}{T}
\label{eq:def_powerenergy}
\end{align}
for a voltage $v(t)$ and current $i(t)$ with arbitrary time dependence. 
Corresponding definitions apply to random $P(t)$, $U(T)$ and $P_{\rmeff}$, which pose no issue provided that the rate of dynamic variation does not exceed that of the electrical fluctuation for the field ($\sqrt{\lambda^\prime_2} / \omega <  1$).
This is the case for classical mechanical stirring, 
viz., $\sqrt{\lambda^\prime_2}/\omega = 9.4 \times 10^{-9}$ in the present experiment, but may approach $1$ more closely in rapid electronic stirring. The power associated with excursion intervals is then obtained by dividing the excess energy by the corresponding excursion length. 
For rapid variations ($\sqrt{\lambda^\prime_2} / \omega \not \ll 1$), the fluctuation of the field and its time-varying envelope must be taken into account, which is considered next. 

For a time-invariant amplitude (peak voltage) $v$, the electric energy accumulated across $[0,T]$ is
\bea
u(T) =\int^T_0 [v \cos(\omega t)]^2 \rmd t = \frac{v^2}{2} \left ( T + \frac{\sin(2\omega T)}{2\omega} \right ).
\label{eq:energy_LTI}
\eea
The effective power $p_{\rmeff}$ with reference to $[0,T]$ follows 
%from (\ref{eq:def_powerenergy}) 
as
\bea
p_{\rmeff} = \left ( \frac{v}{\sqrt{2}} \right )^2 \left [ 1 + {\sinc} (2\omega T) \right ] \stackrel{\small \omega T\rightarrow \infty}{\longrightarrow} \left ( \frac{v}{\sqrt{2}} \right )^2
\eea
whence $v_{\rmeff}=v/\sqrt{2}$ in the quasi-stationary low-frequency (LF) limit is retrieved.
In analogy with (\ref{eq:def_norm}), the corresponding random quantities are further normalized as
\begin{align}
{P^{(+)}_{(\rmeff)}}^\prime \stackrel{\Delta}{=} \frac{P^{(+)}_{(\rmeff)}}{\sigma^2_V},~~
{U^{(+)}_{(\rmeff)}}^\prime \stackrel{\Delta}{=} \frac{U^{(+)}_{(\rmeff)}}{\sigma^2_V},~~
V^\prime_{(\rmeff)} \stackrel{\Delta}{=} \frac{V_{(\rmeff)}}{\sigma_V}
.
\end{align}

As is well known (cf., e.g., \cite[sec. 6.2.2]{levin1966}), the definition of the {\em real\/} envelope $V(t)$ -- as used in defining the effective voltage and effective power -- holds regardless of the bandwidth of the process. Therefore, the following exposition is valid regardless of the value of $\sqrt{(\lambda^\prime_2)_i}/\omega_j$ for the spectral components of the motion vs. excitation.

For $u^\prime_{\rmth} \gg 1$, the SKM (\ref{eq:Slepian_up}) holds and the duration is
\bea
L^+(u^\prime_{\rmth}) = 2 (t^+_{\rmmax} - t^+ ) = \frac{2Z}{\omega\kappa^+}
\eea
in which the dimensionless parameter $\kappa^+$ is
\bea
\kappa^+ \stackrel{\Delta}{=} m \sqrt{u^\prime_{\rmth}} = \sqrt{\lambda^\prime_2u^\prime_{\rmth}}/{\omega}
\eea
and where $m$ is the HM index (\ref{eq:HMindex}).
The accumulated energy follows from  (\ref{eq:energy_LTI}) by integrating $P(t)=V^2(t)/R$ across $T=L^+$, with $V^2(t)$ given by (\ref{eq:Slepian_up}) and referencing to a unit resistance $R$, yielding 
\begin{align}
U^\prime(Z)
&= \left ( \frac{u^\prime_{\rmth}}{\omega \kappa^+} - \frac{\kappa^+}{4 \omega} \right ) Z + \frac{2\,Z^3}{3 \omega \kappa^+} - \frac{\kappa^+ Z}{4 \omega} \cos \left ( \frac{4 Z}{\kappa^+} \right )\nonumber\\
&~  + \left ( \frac{u^\prime_{\rmth}}{4\omega} + \frac{{\kappa^+}^2}{8\omega} \right ) \sin \left ( \frac{4 Z}{{\kappa^+}} \right ).
\label{eq:Upls_tot}
\end{align}
With $Z$ taking a value $z_i$, this results in $u^\prime(z_i)$ and $p^\prime_i=u^\prime(z_i)/\ell^+_i$.
Finally, the effective power across $L^+$ follows 
as
\begin{align}
P^\prime_{\rmeff}(Z) 
&= \frac{U^\prime(Z)}{T} = \frac{\omega{\kappa^+}}{2Z} U^\prime(Z) \nonumber\\
&= \frac{u^\prime_{\rmth}}{2} - \frac{{\kappa^+}^2}{8} + \frac{Z^2}{3} - \frac{{\kappa^+}^2}{8} \cos \left ( \frac{4 Z}{{\kappa^+}} \right ) \nonumber\\
&~ + \left ( \frac{u^\prime_{\rmth}}{2} + \frac{{\kappa^+}^2}{4} \right ) \sinc \left ( \frac{4 Z}{{\kappa^+}} \right )
.
\label{eq:Ppls_final}
\end{align}
In (\ref{eq:Upls_tot}) and (\ref{eq:Ppls_final}), the harmonic terms contain amplitudes that depend on  $\kappa^+$, while the phase $4Z/\kappa^+$ is Rayleigh distributed, i.e., nonuniform. The ratio $P^\prime_{\rmeff} / (u^\prime_{\rmth} / 2)$ measures the relative contribution of the EME dynamics to the power in the excursion.

For arbitrary $u^\prime_{\rmth}$, two limiting cases can be identified:
\begin{itemize}
\item quasi-static LF excitation or exceedingly ``high'' EME velocity: this implies $1 / {\kappa^+} \rightarrow 0$, i.e., $\sqrt{\lambda^\prime_2}/\omega \rightarrow + \infty$ for finite $u^\prime_{\rmth}$ and $Z$. In this case, a series expansion of (\ref{eq:Ppls_final}) leads to
\begin{align}
P^\prime_{\rmeff}(Z) \simeq u^\prime_{\rmth} + \frac{2\,Z^2}{3} - \frac{4 \, Z^4}{5\, (\kappa^+)^2} + {\cal O} \left [ \frac{Z^6}{(\kappa^+)^4} \right ];
\end{align}
\item quasi-optical HF excitation or quasi-statically ``low'' EME velocity: this corresponds to $\kappa^+ \rightarrow 0$. In this classic regime, it follows that
\begin{align}
P^\prime_{\rmeff}(Z) \rightarrow \frac{1}{2} \left ( u^\prime_{\rmth} + \frac{2\,Z^2}{3} \right ).
\end{align}
In this case, $P^+_{\rmeff}$ maps to half of the area of the parabolic cap $A^+/2$, for which the PDF (\ref{eq:PDF_Apls}) applies. Together with the area $A_{\rmth}$ and the root mean square factor $1/2$ for effective scaling, this yields $P^\prime_{\rmeff}$ as expected. 
\end{itemize}

The PDF $f_{P^\prime_{\rmeff}}(p^\prime_{\rmeff})$ can be obtained from the inverse variate transformation $Z\rightarrow P^\prime_{\rmeff}$ in (\ref{eq:Ppls_final}). Alternatively, and more conveniently, this PDF is also obtained by Monte Carlo (MC) simulation through generating $N$ Rayleigh distributed values of $Z$. 
Fig. \ref{fig:CCDF_Ppls} shows the thus obtained complementary CDF (CCDF) $1-F_{P^\prime_{\rmeff}}(p^\prime_{\rmeff})$ of Cartesian power ($n=2$) for $u^\prime_{\rmth}=10$ at selected values of $m$, based on $N=10^5$ independent replicas. 
The tail dependence for $m \ll 1$ indicates that quasi-static $P^\prime_{\rmeff}$ as an area-to-length ratio also follows a GPD distribution with $\gamma=0$.
%**************FIGURE 10****************
\begin{figure}[!ht] 
\begin{center} \begin{tabular}{c}
\vspace{-3cm}
\hspace{-0.8cm}
%FIGURE 20 IN excesspower_upcross_pdf_MC_v3.m
\includegraphics[scale=0.5]{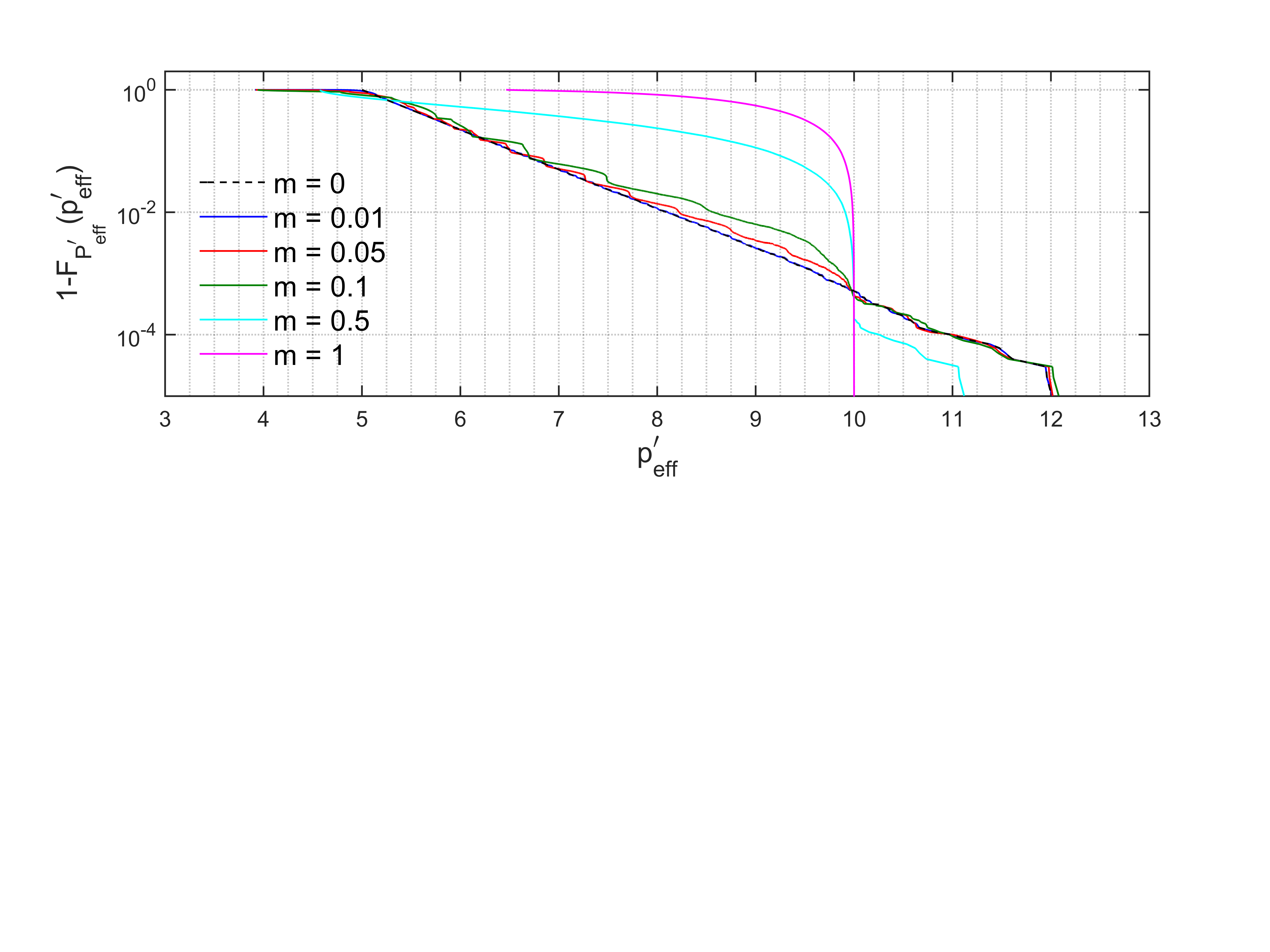}\ \\
\end{tabular} \end{center}
{\small 
\caption{\label{fig:CCDF_Ppls}
\small CCDF of Cartesian $P^\prime_{\rmeff}$ for $u^\prime_{\rmth}=10$ at selected values of HM index $m$, based on $N=10^5$ MC replicas.
}
}
\end{figure}
%**************FIGURE 10****************

The mean of $P^\prime_{\rmeff}$ is obtained after calculation as
\begin{align}
\langle P^\prime_{\rmeff} \rangle 
&= \int^{+\infty}_0 P^\prime_{\rmeff}(z) f_Z(z) \rmd z
=\frac{2}{3} + \frac{u^\prime_{\rmth}}{2} - \frac{{\kappa^+}^2}{4} \nonumber\\
&~~ + \left [ \frac{\sqrt{2}}{2} \left ( 1 + \frac{u^\prime_{\rmth}}{4} \right ) {\kappa^+} + \frac{{\kappa^+}^3}{8 \sqrt{2}} \right ] 
D\left ( \frac{2\sqrt{2}}{{\kappa^+}} \right )
\label{eq:avgP_gen}
\end{align}
where $D(\cdot)$ is Dawson's integral \cite{abra1972}, defined by
\begin{align}
D(x) \stackrel{\Delta}{=} \int^x_0 \exp(t^2-x^2) \rmd t = \frac{\sqrt{\pi}}{2} \exp (-x^2) {\rmerfi}(x) 
\end{align}
in which ${\rmerfi}(x) \stackrel{\Delta}{=} (-\rmj) {\rmerf}(\rmj x)$, with ${\rmerf}(\cdot)$ denoting the (real) error function.
Asymptotic expansions up to second order in ${\kappa^+}$ follow as
\begin{align}
\langle P^\prime_{\rmeff} \rangle = 
\left \{
\begin{array}{ll}
\frac{2}{3} + \frac{u^\prime_{\rmth}}{2} + \frac{{\kappa^+}^2}{4} (3 + u^\prime_{\rmth}) + \ldots, & \kappa^+ \ll 1 \\
\\
\frac{4}{3} + u^\prime_{\rmth} - \frac{8}{3 {\kappa^+}^2} \left( \frac{2}{5} + u^\prime_{\rmth} \right ) + \ldots , & {\kappa^+} \gg 1
\end{array}
\right.
\end{align}
confirming that $\langle P^\prime_{\rmeff}({\kappa^+}\rightarrow +\infty) \rangle = 2 \langle P^\prime_{\rmeff}({\kappa^+}\rightarrow 0) \rangle$.
It follows that for $u^\prime_\rmth \rightarrow +\infty$, the ratio of the average above-threshold excess power $\langle P^+_{\rmeff} \rangle$ to the average below-threshold power $\langle P_{{\rmeff}, \rmth} \rangle$ for ${\kappa^+} \ll 1$ equals this ratio for ${\kappa^+} \gg 1$, to second order in $\kappa^+$, viz., $(2/3)/(u^\prime_{\rmth}/2) = 4/(3 u^\prime_{\rmth})$. Specifically,
\begin{align}
\frac{\langle P^+_{\rmeff} \rangle}{\langle P_{{\rmeff},\rmth} \rangle} = 
\left \{
\begin{array}{ll}
\frac{4}{3 u^\prime_{\rmth}} \left ( 1 + \frac{5{\kappa^+}^2}{8} + \ldots \right ), & \kappa^+ \ll 1\\
\\
\frac{4}{3u^\prime_{\rmth}} \left ( 1 + \frac{28}{15{\kappa^+}^2} + \ldots \right ), & \kappa^+ \gg 1.
\end{array}
\right.
\end{align}
For arbitrary ${\kappa^+}$, this ratio is frequency dependent, as is also apparent from (\ref{eq:avgP_gen}).
Since $P^+_{\rmeff}$ arises as the difference of effective powers $P_{\rmeff}$ and $P_{\rmeff,\rmth}$ from ground level up, $P^+_{\rmeff}$ represents an effective power quantity as well. 

The second moment $\langle {P^\prime_{\rmeff}}^2 \rangle\stackrel{\Delta}{=}\int^{\infty}_0 [P^\prime_{\rmeff}(z)]^2 f_Z(z)\rmd z$ follows by a similar integration. Combining with (\ref{eq:avgP_gen}) yields the variance of $P^\prime_{\rmeff}$ as 
\begin{align}
&\, \sigma^2_{P^\prime_{\rmeff}} =
\frac{16}{9} + \frac{u^\prime_{\rmth}}{3} + \frac{{\kappa^+}^2}{6} - \sqrt{2} \left [ 
\frac{16}{3{\kappa^+}} + \frac{{\kappa^+}}{3} - \frac{{\kappa^+}^3}{12} - \frac{{\kappa^+}^5}{64} \right . \nonumber\\
&~ \left. + 
\left ( \frac{4}{3{\kappa^+}} + \frac{{\kappa^+}}{12} - \frac{{\kappa^+}^3}{32} \right ) 
u^\prime_{\rmth} \right ] 
D\left ( \frac{2\sqrt{2}}{{\kappa^+}} \right )
\nonumber\\
&~ - \sqrt{2} \left ( \frac{{\kappa^+}^3}{16} + \frac{{\kappa^+}^5}{128} +\frac{{\kappa^+}^3 u^\prime_{\rmth}}{64} \right ) D\left ( \frac{4\sqrt{2}}{{\kappa^+}} \right )
\nonumber\\
&~ - \frac{1}{2} \left [ \left ( 1 + \frac{u^\prime_{\rmth}}{4} \right ) {\kappa^+} + \frac{{\kappa^+}^3}{8} \right ]^2 D^2\left ( \frac{2\sqrt{2}}{{\kappa^+}} \right ) \nonumber\\
&~ + \left( \frac{{\kappa^+}^4}{16} + \frac{{u^{\prime}_{\rmth}}^2}{4} + \frac{u^\prime_{\rmth} {\kappa^+}^2}{4} \right ) \/_2F_2\left (1,1;\frac{3}{2},2;-\frac{32}{{\kappa^+}^2} \right )
\label{eq:sigmasqP_gen} 
\end{align}
where $\/_2F_2\left (\cdot,\cdot;\cdot,\cdot; y \right )$ is a generalized hypergeometric series.

Fig. \ref{fig:excesspower_avgstdcffvar} shows $\langle P^\prime_{\rmeff} (m)\rangle/u^\prime_{\rmth}$ and $\sigma_{P^\prime_{\rmeff}}(m)/u^\prime_{\rmth}$ together with $\nu_{P^\prime_{\rmeff}}(m) = \sigma_{P^\prime_{\rmeff}}(m)/\langle {P^\prime_{\rmeff}} (m) \rangle$, for selected  values of $u^\prime_{\rmth}$. 
These characteristics were calculated from (\ref{eq:avgP_gen}) and (\ref{eq:sigmasqP_gen}), and subsequently validated by MC simulation of (\ref{eq:Ppls_final}), thus also validating the CCDFs in Fig. \ref{fig:CCDF_Ppls} in the process. 
The dependencies for $m>1$ are hypothetical, shown as dashed curves. 
The increase of the power is more pronounced for increased $u^\prime_{\rmth}$. 
At the same time, the interval with resonance-like variation shifts towards lower $m$, with this transition region showing first relatively large positive, then negative variation compared to the quasi-static values for $m\rightarrow 0$.
For slow dynamics and LF excitation, typical values of the maximum-to-mean value of $P^\prime_{\rmeff}$ are of the order of $u^\prime_{{\rmth}, {\rmmax}} \ll 10$ and indicate that the quasi-static approximation for the mean and standard deviation of the power is excellent, i.e., valid for non-extreme speeds. 
%**************FIGURE 11****************
\begin{figure}[!ht] 
\begin{center} \begin{tabular}{c}
\vspace{-3.8cm}
\hspace{-0.9cm}
%FIG 1 IN EXCESSPOWERSTATS_V3.M
(\includegraphics[scale=0.48]{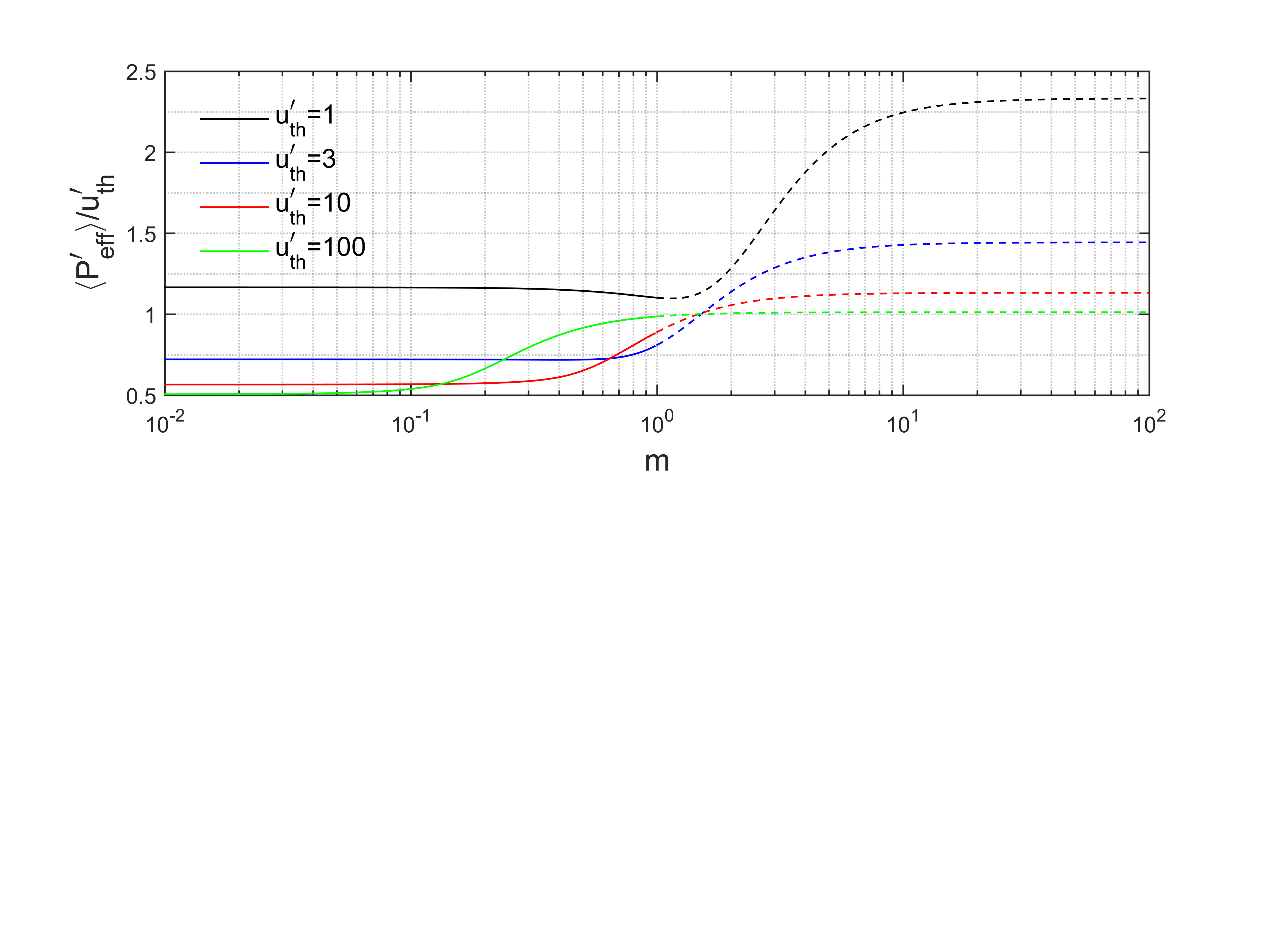}\ \\
(a)\\
\vspace{-3.8cm}
\hspace{-0.9cm}
%FIG 2 IN EXCESSPOWERSTATS_V3.M
\includegraphics[scale=0.48]{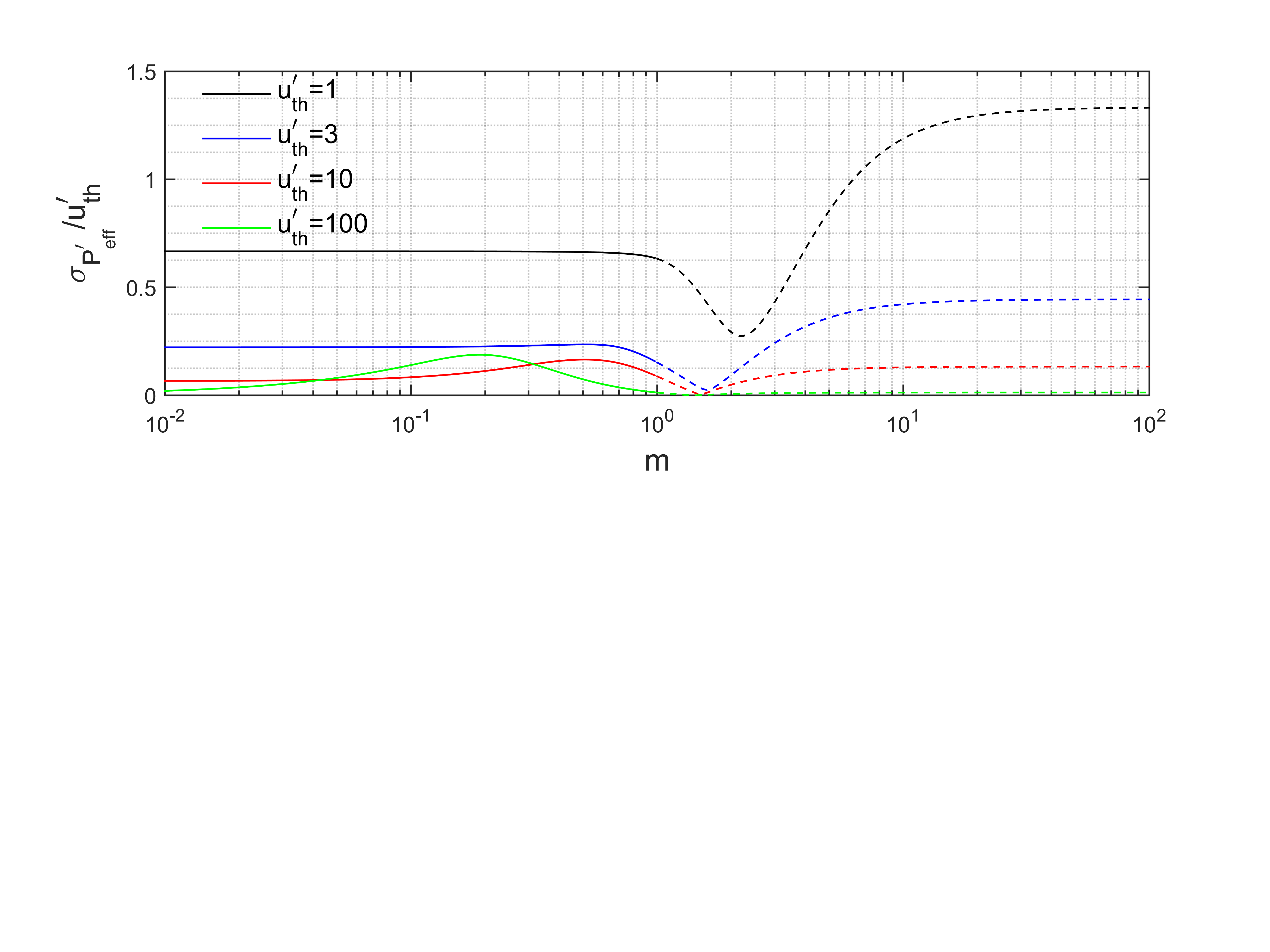}\ \\
(b)\\
\vspace{-3.7cm}
\hspace{-0.9cm}
%FIG 3 IN EXCESSPOWERSTATS_V3.M
\includegraphics[scale=0.48]{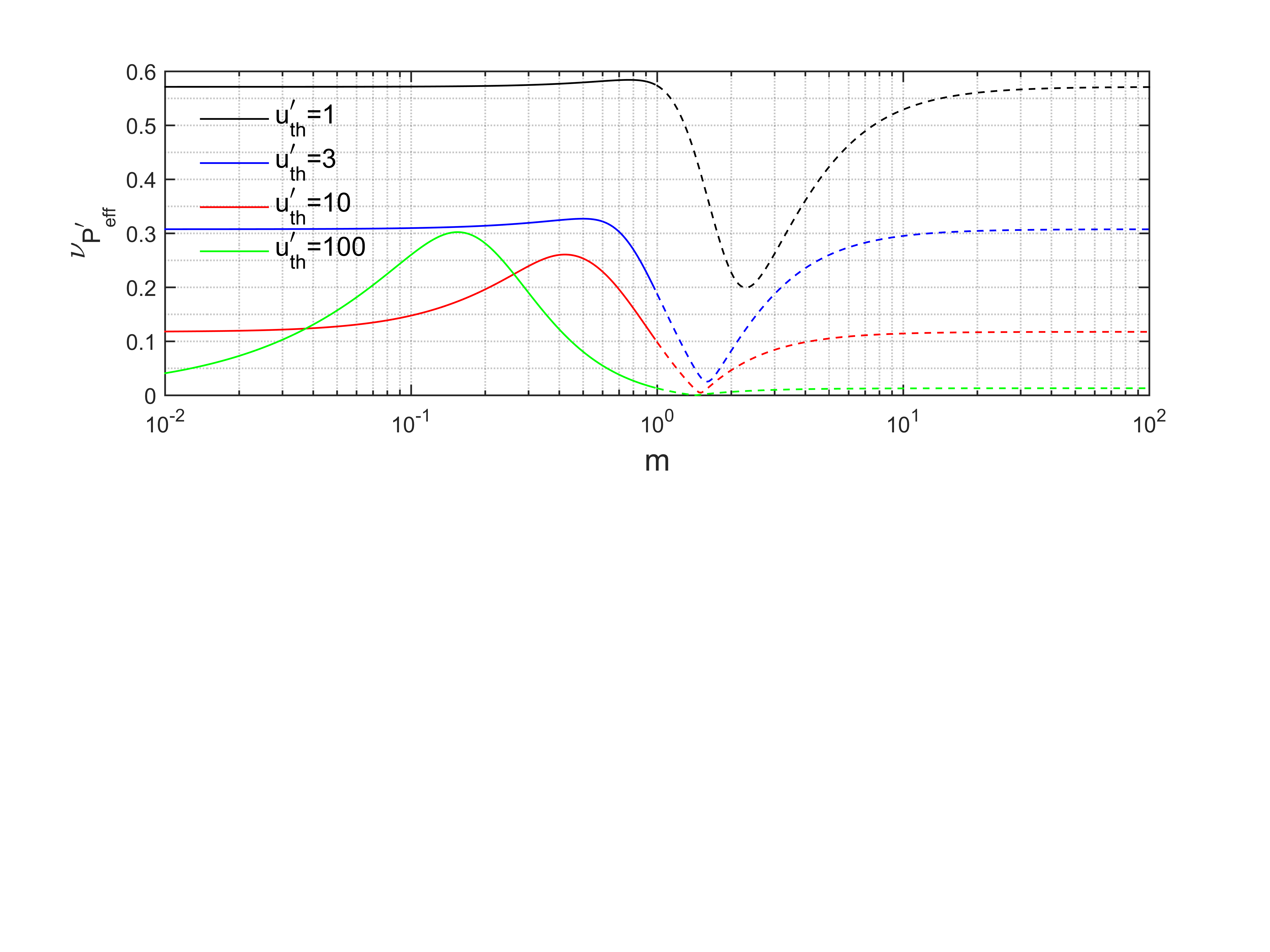}\ \\
(c)\\
\end{tabular} \end{center}
{\small 
\caption{\label{fig:excesspower_avgstdcffvar}
\small
(a) Normalized mean,
(b) normalized standard deviation, and
(c) coefficient of variation of dynamic power $P^\prime_{eff}$ as a function of modulation index $m$ for selected values of $u^\prime_{\rmth}$.
}
}
\end{figure}
%**************FIGURE 11****************

\section{Conclusion}
EM fields can become modulated at the source, using traditional electrical modulation techniques, or through propagation effects in a dynamic EME. In this work, the relative contribution of both mechanisms in the power of the received field above high threshold levels was investigated, based on a second-order parabolic SKM conditioned on upcrossings for the excursions of intensity. 
Asymptotic and regressed PDFs of the excursion height and area were obtained as (\ref{eq:exp_height_chisq}), (\ref{eq:PDF_Upls_firstregress}), (\ref{eq:PDF_Apls}) and (\ref{eq:PDF_Apls_firstregress}), representative of quasi-static field intensity and energy. 
For dynamic EMEs, the mean (\ref{eq:avgP_gen}) and variance (\ref{eq:sigmasqP_gen}) of the effective excess power (\ref{eq:Ppls_final}) were obtained, whose CCDF was calculated numerically (Fig. \ref{fig:CCDF_Ppls}).

For sufficiently high threshold crossing levels, the decomposition provided by a SKM permits the extraction of the quasi-deterministic short-term contribution to the field, which dominates during level excursions. This greatly simplifies its representation, in particular for the PDF of the regressed early energy. 
For the field intensity, an SKM 
yields a more accurate estimation of local and global maximum values with a reduced uncertainty (Figs. \ref{fig:Slepian_musigma} and \ref{fig:Slepian_Tektronix_parabolae}), compared with a traditional statistical characterization based on unconditioned sampled fields (i.e., without thresholding). The latter is retrieved as the long-term asymptotic in the SKM. 
For relatively low thresholds, the interdependence between lengths of consecutive positive and negative excursions \cite{sire2007}  and the partial correlation between excursion height and length make accurate modelling for such thresholds more complicated.
  
Compared to other approaches to conditional statistical characterization, e.g., based on prior PDFs in Bayesian inference \cite{caro2017}, the SKM only uses timings of upcrossings (or another choice of marks) together with the Rayleigh PDF (\ref{eq:fZ}) for the slope, thus simplifying the modelling. 
Using a prior PDF for the bandwidth--level product $\sqrt{\lambda^\prime_2 u^\prime_{\rmth}}$ as parameter, the likelihood obtained in the SKM can be used to yield posterior PDFs, in particular for sampling distributions for short intervals.

\end{document}